\documentclass[narrow]{elsart1p}

\usepackage{natbib}
\usepackage{epsfig}
\usepackage{amssymb}
\journal{New Astronomy}

\newcommand{\ga}{\gtrsim}
\newcommand{\la}{\lesssim}
\newcommand{\bl}[1]{\mbox{\boldmath$ #1 $}}
\newcommand{\vnp}{\bl{v}_{\rm n}}
\newcommand{\vip}{\bl{v}_{\rm i}}
\newcommand{\cs}{c_{\rm s}}
\newcommand{\csq}{c_{\rm s}^2}
\newcommand{\ceffsq}{C^2_{\rm eff}}
\newcommand{\zhat}{\mbox{\boldmath$ \hat{z}$}}

\newcommand{\beq}{\begin{equation}}
\newcommand{\eeq}{\end{equation}}
\newcommand{\barray}{\begin{eqnarray}}
\newcommand{\earray}{\end{eqnarray}}

\newcommand{\Msun}{M_{\odot}}
\newcommand{\ul}{\underline{\hspace{20pt}}}
\newcommand{\bfig}{\begin{figure}}
\newcommand{\efig}{\end{figure}}
\newcommand{\FT}{{\cal F}}
\newcommand{\FTinv}{{\cal F}^{-1}}

\newcommand{\cmc}{~{\rm cm}^{-3}}
\newcommand{\cms}{~{\rm cm}^{-2}}

\newcommand{\kms}{~{\rm km~s}^{-1}}

\newcommand{\pc}{~{\rm pc}}

\newcommand{\K}{~{\rm K}}
\newcommand{\muG}{~\mu{\rm G}}
\newcommand{\yr}{~{\rm yr}}
\newcommand{\Myr}{~{\rm Myr}}
\newcommand{\AU}{~{\rm AU}}
\newcommand{\mH}{m_{\rm H}}
\newcommand{\Htwo}{{\rm H}_{2}}
\newcommand{\mn}{m_{\rm n}}
\newcommand{\mi}{m_{\rm i}}
\newcommand{\Pext}{P_{\rm ext}}
\newcommand{\Pexttil}{\tilde{P}_{\rm ext}}
\newcommand{\mui}{\mu_0}

\newcommand{\tnii}{\tau_{\rm ni,0}}
\newcommand{\tniitil}{\tilde{\tau}_{\rm ni,0}}
\newcommand{\rhon}{\rho_{\rm n}}
\newcommand{\rhoni}{\rho_{\rm n,0}}
\newcommand{\nn}{n_{\rm n}}

\newcommand{\nni}{n_{\rm n,0}}
\newcommand{\nion}{n_{\rm i}}
\newcommand{\nii}{n_{\rm i,0}}
\newcommand{\sign}{\sigma_{\rm n}}

\newcommand{\signi}{\sigma_{\rm n,0}}
\newcommand{\signmax}{\sigma_{\rm n,max}}
\newcommand{\Nni}{N_{\rm n,0}}
\newcommand{\Beq}{B_{z,\rm eq}}

\newcommand{\Beqi}{B_{z,\rm eq,0}}
\newcommand{\Bref}{B_{\rm ref}}
\newcommand{\Breftil}{\tilde{B}_{\rm ref}}

\newcommand{\vnmax}{|v_{\rm n}|_{\rm max}}
\newcommand{\vimax}{|v_{\rm i}|_{\rm max}}

\newcommand{\lammax}{\lambda_{\rm T,m}}
\newcommand{\lamgm}{\lambda_{\rm g,m}}
\newcommand{\mgm}{M_{\rm g,m}}
\newcommand{\taugm}{\tau_{\rm g,m}}
\newcommand{\kion}{k_{\rm i}}
\newcommand{\xion}{x_{\rm i}}
\newcommand{\lamavg}{\langle \lambda \rangle}
\newcommand{\Mavg}{\langle M \rangle}
\newcommand{\savg}{\langle s \rangle}
\newcommand{\axisravg}{\langle b/a \rangle}
\newcommand{\Zavg}{\langle Z \rangle}
\newcommand{\sigin}{\langle \sigma w \rangle_{\rm{i\Htwo}}}
\newcommand{\Psim}{\Psi_{\rm M}}

\begin{document}

\begin{frontmatter}

\title{Nonlinear Evolution of Gravitational Fragmentation Regulated by
Magnetic Fields and Ambipolar Diffusion}

\author[label1]{Shantanu Basu\corauthref{cor}}, 
\corauth[cor]{Corresponding author.}
\ead{basu@astro.uwo.ca}
\author[label2]{Glenn E. Ciolek},
\ead{cioleg@rpi.edu}
and \author[label1]{James Wurster}

\address[label1]{Department of Physics and Astronomy, University of
Western Ontario, London, Ontario N6A~3K7, Canada}
\address[label2]{Department of Physics, Applied Physics, and Astronomy,
Rensselaer Polytechnic Institute, 110 W. 8th Street, Troy, NY 12180,
USA}

\begin{abstract}
We present results from an extensive set of simulations of gravitational
fragmentation in the presence of magnetic fields and ambipolar diffusion.
The thin-sheet approximation is employed, with an ambient magnetic field
that is oriented perpendicular to the plane of the sheet.
Nonlinear development of fragmentation instability leads to substantial
irregular structure and distributions of fragment spacings, fragment masses,
shapes, and velocity patterns in model clouds. We study the effect of 
dimensionless free parameters that characterize the 
initial mass-to-flux ratio, neutral-ion coupling, and external pressure 
associated with the sheet.
The average fragmentation spacing in the nonlinear phase of evolution is in
excellent agreement with the prediction of linear perturbation theory.
Both significantly subcritical {\it and} highly supercritical clouds 
have average fragmentation scales $\lamavg \approx 2 \pi Z_0$, where   
$Z_0$ is the initial half-thickness of the sheet. In contrast, the
qualitatively unique transcritical modes can have $\lamavg$ that 
is at least several times larger. Conversely, fragmentation dominated by
external pressure can yield dense cluster formation with much smaller
values of $\lamavg$. The time scale for nonlinear
growth and runaway collapse of the first core is $\approx 10$ times 
the calculated growth time $\taugm$ of the eigenmode with minimum growth time,
when starting from a uniform background state with small-amplitude
white-noise perturbations. Subcritical and transcritical models 
typically evolve on a significantly longer time scale than the supercritical
models.
Infall motions in the nonlinear fully-developed contracting cores
are subsonic on the core scale in subcritical and transcritical clouds,
but are somewhat supersonic in supercritical clouds. 
Core mass distributions are sharply peaked with a steep decline to large
masses, consistent with the existence of a preferred mass scale for each
unique set of dimensionless free parameters. 
However, a sum total of results for various 
initial mass-to-flux ratios yields a broad distribution reminiscent of observed
core mass distributions.
Core shapes are mostly near-circular in the plane of the sheet for subcritical
clouds, but become progressively more elongated for clouds with increasing
initial mass-to-flux ratio. Field lines above the cloud midplane
remain closest to vertical in the ambipolar-drift driven core formation
in subcritical clouds, and there is increasing amount of magnetic field 
curvature for clouds of increasing mass-to-flux ratio. 
Based on our results, we conclude that fragmentation spacings, 
magnitude of infall motions, core
shapes, and, especially, the curvature of magnetic field 
morphology, may serve as  
indirect observational means of determining a cloud's ambient
mass-to-flux ratio.
\end{abstract}


\end{frontmatter}

\section{Introduction}
\label{s:intro}
\subsection{Molecular Cloud Cores}

Star formation occurs in dense cores within
interstellar molecular clouds.
The existence and properties of dense cores
are well established by studies of molecular spectral line emission 
\citep{Myer83b,Bens,Jiji}, submillimeter dust emission 
\citep{Ward94,Andr,Kirk}, and infrared absorption
\citep{Bacm,Teix,Lada07}. 
The core formation process 
has been the subject of intense theoretical study for the past few 
decades. Ideas range from 
uninhibited gravitational fragmentation instability 
\citep{Jean,Lars85,Lars03},
to ambipolar diffusion in a magnetically supported cloud 
\citep{Mest,Mous78,Shu87},
to a very rapid fragmentation due to pre-existing turbulent flows
\citep{Pado,Kless,Gamm}.
The last two ideas are related to scenarios for the nature of relatively
low-density molecular cloud envelopes, which contain most of the mass 
of molecular clouds \citep[see][]{Gold}. 
In one scenario, they are magnetically-dominated
and evolve due to ambipolar diffusion, a gravitationally-driven redistribution of mass 
amongst magnetic flux tubes \citep{Mous78}. The relatively low
global star formation rate and efficiency in Galactic molecular clouds
\citep[see][]{McKe} is used to argue
that molecular cloud envelopes have a subcritical mass-to-flux ratio 
\citep[see e.g.][]{Shu99,Elme07}. 
Indirect empirical evidence based on the Chandrasekhar-Fermi (CF) 
method and velocity anisotropy measurements
does imply that the low density ($n \sim 100 \cmc$) regions are 
subcritical \citep{Cort,Heye}. The contrasting view is that
the mass-to-flux ratio is supercritical on large scales, and that the
star formation rate and efficiency are governed by supersonic turbulence
\citep{MacL}. This requires either a continual replenishment of rapidly dissipating
turbulence, or a relatively rapid dispersal of the cloud envelopes.
The focus of this paper 
is the fragmentation of embedded and relatively quiescent 
dense regions, and not on the nature of the larger envelopes.
We study the effect of small-amplitude perturbations
on a variety of cloud models with different mass-to-flux ratios, 
degrees of magnetic coupling, and external pressures.
 

For the purpose of this paper, we refer to any individual unit of star
formation, which leads typically to a single or small multiple star system, 
as simply a ``core''. However, astronomers often subdivide this concept
into two observational categories, that of  
``prestellar core'' and ``prestellar condensation''. 
The former term is often used to describe 
the extended (mean density $n \sim 10^5 \cmc$, size $s \sim 0.1$ pc,
spacing $\lambda \sim 0.25$ pc) objects in regions of distributed
star formation like the Taurus molecular cloud, while the latter
term typically describes the more compact ($n \ga 10^6-10^7 \cmc$,
$s \sim 0.02-0.03$ pc, $\lambda \sim 0.03$ pc) objects that are located 
within cluster-forming cores in, for example, Ophiuchus, Serpens,
Perseus, and Orion \citep[see discussion in][]{Ward07}. 
Both the prestellar cores in e.g., Taurus, and the
cluster-forming cores (which harbor multiple condensations) 
in the other clouds occupy only a very small fraction of the total 
volume and mass of their larger, more turbulent, molecular cloud complex
\citep[e.g.,][]{John,Gold}.
A compendium of current data from
spectral line emission, dust emission, and infrared absorption tends to
show that cores exhibit central density concentration
\citep{Ward94,Andr}
subsonic inward motions \citep{Tafa,Will,Lee,Case} that often extend beyond the 
nominal core boundary, near-uniform gas temperatures \citep{Bens}, subsonic 
internal turbulence \citep{Myer83a,Full,Good98}, 
and near-critical magnetic field strengths, when detected by the Zeeman
effect \citep{Crut99,Bour} or inferred by the CF method 
\citep{Crut04}.

The measured subsonic infall motions constitute indirect evidence for a force
that mediates gravity.
The mediative force may be
due to the magnetic field, whose strength in dense regions is very close to 
the critical value for collapse \citep{Crut99}.
If the fragmenting region has a subcritical mass-to-flux 
ratio, then the formation of cores will be regulated by
ambipolar diffusion, i.e. neutral molecules diffusing past ions that
are tied to magnetic fields, and will occur on a time scale that may significantly
exceed the local dynamical time for typical ionization fractions. 
Conversely, if the dense gas is supercritical, then the fragmentation takes 
place on a dynamical time scale. \citet[hereafter BC04]{BC04} studied the nonlinear
development of fragmentation instability in clouds that were initially 
either critical or decidedly supercritical. They found that supercritical
fragmentation is characterized by somewhat supersonic motions on the 
core scale ($\sim 0.1$ pc) while critical fragmentation is characterized by
subsonic inward motions on those scales. The nonlinear fragmentation of 
decidedly subcritical clouds
is also presented in this paper as part of a comprehensive 
parameter study of
the effects of mass-to-flux ratio, initial ionization fraction, and external pressure.
We believe that the three broad categories of subcritical, transcritical, 
and supercritical fragmentation should all occur in the interstellar medium,
and even within separate regions of a single molecular cloud complex. Our
results can help to distinguish which mode of fragmentation is occurring in 
a given observed region.

The linear theory of fragmentation of a partially ionized, magnetic thin sheet
has been presented by \citet[hereafter CB06]{CB06}. 
An important quantity is the dimensionless critical mass-to-flux ratio, 
$\mu = 2 \pi G^{1/2} \sigma/B$, where $\sigma$ is the column density of a 
sheet and $B$ is the strength of the magnetic field that is oriented 
perpendicular to the plane of the sheet. In the limit of flux-freezing,
fragmentation can occur only if $\mu > 1$, i.e. the sheet is supercritical. 
Conversely, gravitationally-driven fragmentation instability cannot occur at all 
if $\mu < 1$, i.e. the sheet is subcritical.
CB06 show that the inclusion of ambipolar drift in such a model means that 
fragmentation can occur for {\it all mass-to-flux ratios}, but on widely 
varying time scales and length scales. 
A very important result of CB06
is their Fig. 2 (see also Fig. 1 in this paper), which demonstrates that transcritical ($\mu \approx 1$) fragmentation
has a preferred scale that can be many times larger
than $\lammax$, the wavelength of maximum growth rate in the thermal
(i.e. nonmagnetic) limit. The latter
is actually the preferred scale for both highly supercritical {\it and} 
highly subcritical clouds. This is because gravitational
instability in subcritical clouds develops by ambipolar drift of neutrals 
past near-stationary
magnetic field lines, and occurs on the ambipolar diffusion
time scale rather than the dynamical time scale \citep[see also][]{Lang78,Mous78}.
The ``resonant'' transcritical modes with preferred scale $\lamgm \gg \lammax$ 
occur due to a combination of ambipolar drift and field-line dragging;
the latter leads to magnetic restoring forces that can stabilize the
perturbations unless they are of the required large size.
The time scale for ambipolar-diffusion mediated gravitational instability
is also presented extensively in CB06. They showed that highly
supercritical clouds undergo gravitational instability on the 
dynamical time $t_{\rm d} 
\approx Z/\cs$, where $Z$ is the half-thickness (effectively the scale height)
of the sheet and $\cs$ is the isothermal sound speed.
On the other hand,
highly subcritical clouds undergo instability on the
quasistatic ambipolar diffusion
time $\tau_{\rm AD} \approx 10\,Z/\cs$ for typical ionization
fraction. Transcritical clouds undergo a hybrid instability on an
intermediate time scale -- see also \citet{Zwei} for a 
similar result. Also, we note that the inclusion of nonlinear fluctuations 
can reduce the ambipolar diffusion time under certain circumstances
\citep{Fatu,Zwei02}; we do not model such effects in this paper. 
 
Our nonlinear simulations represent a significant extension of the
parameter space of models from that presented by BC04. It similarly 
extends the parameter space studied by \cite{Inde}, who presented
nonaxisymmetric evolution of an infinitesimally thin subcritical sheet, 
including the effects of magnetic tension but ignoring magnetic pressure.
Recent fully three-dimensional simulations by \citet{Kudo07}, including 
magnetic fields and ambipolar diffusion, have confirmed the basic results
of BC04. However, that paper presents three representative models, and an extensive
parameter study remains computationally out of reach.
Here, in addition to a parameter study, we carry out large numbers of
simulations for {\it each} unique set of parameters. This is done 
in order to compile statistics on core spacings, mass distributions, and shapes.
All simulations end at the time of the runaway central collapse of the first 
core, hence we are compiling information about the early phases of star 
formation in a molecular cloud. The subsequent history of the molecular
cloud, after it has been stirred up by the initial star formation, remains 
to be determined.
This paper represents the product of a total of over
700 separate simulations for models with 14 distinct sets of dimensionless
parameters.

An alternate approach to modeling fragmentation is  
to input turbulence directly into the dense thin-sheet
model \citep[e.g.,][]{Li04,Naka05}, rather than relegating its presence
to an unmodeled envelope. This results in highly supersonic 
motions within the dense sheet itself, and more rapid formation of
cores than in our models. We do not pursue that approach
in the present study but leave it open to future investigation.

\subsection{Relation to Global Cloud Structure}
Our simulations represent an intermediate approach between attempting a
global model of large-scale molecular
cloud structure and of modeling the interior collapse of individual
cores. Large-scale models need to account for the overall
structure and self-gravity of molecular clouds and cannot
be realistically studied using a periodic-box model. 
Truly global three-dimensional models are computationally expensive
and still rarely attempted. Smoothed particle hydrodynamics (SPH)
techniques have been used successfully to model entire
cluster-forming regions \citep{Bate,Bonn}, and can explain some
important features of star formation like mass segregation, binary
fraction, and the initial mass function. These models are not fully global
in that they do not include the effect of the molecular cloud envelopes.
They also do not include the effect of magnetic
fields or feedback from outflows, so they cannot address the observed 
low global star formation efficiency (SFE), $1-5$\% \citep{Lada03}. 
However, the latest version of such models \citep{Pric} does include 
flux-frozen magnetic fields and yields somewhat lower SFE's than
the non-magnetic models.
Another three-dimensional approach which is fully global and based on
finite-difference or SPH techniques is modeling the formation 
of a molecular cloud from large-scale supersonic gas flows and the early evolution
of the cloud \citep{Vazq06,Vazq07,Heit,Henn}.
These models yield an 
initially flattened cloud whose subsequent evolution is affected by
several instabilities (thermal instability, thin-shell instability,
and Kelvin-Helmholtz instability). The overall structure evolves away from
a sheet-like configuration, but individual segments may be treated as such.
Of the above mentioned studies, only the work of \citet{Henn} includes the 
effect of magnetic fields, which may work to suppress some of the
instabilities. It is not clear whether this is an important factor in 
that model. However, the smaller-scale simulation of
\citet{Naka08} does show the formation of a sheet-like structure due to
the presence of a dynamically important magnetic field.  
Amongst our models, the cases with high bounding pressure may be
most appropriate for the scenario of cloud formation due to colliding flows.
Our low external pressure models are more appropriate to the 
alternate scenario where the flows that form a molecular cloud 
are driven by gravity and/or channeled by a large scale magnetic
field, e.g. by the magneto-Jeans instability \citep{Kim} or the 
Parker instability \citep{Park}.

Despite the various theoretical emphases on flattened structure, we note
that observations of molecular clouds reveal complex morphologies with
projected shapes that may not look like globally flattened structures. 
Molecular clouds have been described variously 
in the literature as stratified objects supported by
internally generated turbulence \citep{McKe99}, or as 
fractal objects \citep{Elme96} in which the internal
pressure is not as relevant. Our local sheet 
model may be applicable to the dense subregions of the clouds (where
stars actually form in weak or rich clusters) in either
scenario. As an example from the first scenario described above, 
one-dimensional  
global models \citep{Kudo03,Kudo06,Foli} of molecular clouds 
reveal that internally-driven turbulence 
yields large-amplitude motions in lower-density envelopes, 
while retaining transonic motions in embedded dense regions; the
fragmentation of the latter may be described by our models. 
From an observational point of view, we may apply our models to  
dense star forming regions such as L1495 and HCl 2 in the Taurus
molecular cloud \citep[see][]{Gold}, the L1688 cluster-forming core
in Ophiuchus \citep{Mott}, or better yet to the Pipe Nebula
\citep{Muen}, which represents an even earlier stage of evolution,
with a large number of relatively
quiescent prestellar cores that are found
in a dense elongated region (the ``stem'' of the Pipe). 
This region may represent the best available laboratory
for the study of the early stage of star formation, before 
feedback from star formation has significantly modified a
cloud's internal sructure and motions. There is
also evidence that the stem of the Pipe Nebula is flattened along the direction
of the mean magnetic field \citep{Alve}. This property is similar
to the better established result for the elongated structures in
Taurus \citep{Good90,Gold}.

%




\section{Physical Model}
\label{s:pm}

We consider the evolution of weakly ionized, magnetic
interstellar molecular clouds. The clouds are isothermal, 
having a temperature $T$. As presented in CB06, 
we model a cloud as a planar sheet or layer of infinite extent 
in the $x$- and $y$- directions of a Cartesian coordinate
system ($x$, $y$, $z$). At each time $t$ the sheet has a local 
vertical half-thickness $Z(x,y,t)$. We take our model clouds 
to be {\em thin}: by this we mean that for any physical quantity 
$f(x,y,z,t)$ the condition $f/\nabla_p f \gg Z$ is always satisfied, 
where $\nabla_p \equiv \hat{\bl{x}}\partial/\partial x
+ \hat{\bl{y}}\partial/\partial y$ is the planar gradient 
operator. The magnetic field that threads a cloud has
the form
\beq
\bl{B}(x,y,z,t)= 
\left\{ 
\begin{array}{l}
\Beq(x,y,t)\zhat \hspace{9.3em} \mbox{for $|z| \leq Z(x,y,t)$}, \\
B_{z}(x,y,z,t)\zhat \\ 
+ B_{x}(x,y,z,t)\hat{\bl{x}}
+ B_{y}(x,y,z,t)\hat{\bl{y}} ~~~\mbox{for $|z| > Z(x,y,t)$},  
\end{array}
\right.
\eeq
where $\Beq$ is the vertical magnetic field strength in the equatorial
plane. For $|z| \rightarrow \infty$, $\bl{B} \rightarrow \Bref \zhat$,
where $\Bref$ is a uniform, constant reference magnetic field
very far away from the sheet.
The magnetic field components above the sheet can be determined from $\Beq(x,y)$
at any time using the divergence-free nature
of the magnetic field and the current-free approximation above the sheet
(see CB06 for details).
 
Some simplification is obtained by integrating the physical 
system of equations governing the evolution (conservation of 
mass and momentum, Maxwell's equations, etc.) of a model cloud along 
the vertical axis from $z=-Z(x,y)$ to $z=+Z(x,y)$. In doing so, 
a ``one-zone approximation" is used, in which the density and 
the $x$- and $y$- components of the neutral and ion velocities, 
as well as the $x$- and $y$- components of the gravitational field, 
are taken to be independent of height within the sheet. 
The volume density is calculated from the vertical pressure balance
equation
\beq
\rho \csq = \frac{\pi}{2}G \sign^2 + \Pext + \frac{B_x^2+B_y^2}{8\pi},
\eeq 
where $\Pext$ is the external pressure on the sheet and $B_x$ and 
$B_y$ represent the values at the top surface of the sheet, $z=+Z$.
This simplification is commonly referred to as the ``thin-sheet 
approximation"; the motivation for and the physical reasonability
of it is discussed at length in Section~2 of CB06.
It is the nonaxisymmetric extension of the 
axisymmetric thin-sheet models used to study ambipolar diffusion
and gravitational collapse in magnetic interstellar clouds, as 
originally developed by \citet{CM93} and \citet{BM94}. 

\subsection{Basic Equations}
\label{s:be}

We solve normalized versions of the magnetic thin-sheet
equations as justified in CB06. The unit of velocity is taken
to be $\cs$, the column density unit is $\signi$, 
and the unit of acceleration is $2 \pi G \signi$, equal to the 
magnitude of vertical acceleration above the sheet. Therefore, 
the time unit is $t_0 = \cs/2\pi G \signi$, and the length unit is 
$L_0= \csq/2 \pi G \signi$. From this system we can also construct 
a unit of magnetic field strength, $B_0 = 2 \pi G^{1/2} \signi$. 
The unit of mass is $M_0 = \cs^4/(4\pi^2G^2\,\signi)$.
Here, $\signi$ is the uniform 
neutral column density of the background state, and $G$ is the gravitational
constant.
With these normalizations, the equations used to determine the
evolution of a model cloud are
\barray
\label{cont}
 \frac{{\partial \sign }}{{\partial t}} & = & - \nabla_p  \cdot \left( \sign \, \vnp \right), \\
\label{mom}
 \frac{\partial}{\partial t}(\sign \vnp)  & = & - \nabla_p \cdot (\sign \vnp \vnp) +  \bl{F}_{\rm T}
+ \bl{F}_{\rm M} + \sign \bl{g}_p, \\
\label{induct}
 \frac{{\partial \Beq }}{{\partial t}} & = & - \nabla_p  \cdot \left( \Beq \, \vip \right) , \\
\label{Ftherm}
\bl{F}_{\rm T} & = & - \ceffsq \nabla_p \sign                   ,\\
\label{Fmag}
\bl{F}_{\rm M} & = & \Beq \, ( \bl{B}_p - Z\, \nabla_p \Beq ) + {\cal O}(\nabla_p Z), \\
\label{vieq}
\vip & = & \vnp + \frac{\tniitil}{\sign}\left(\frac{\rhoni}{\rhon}\right)^{\kion} \bl{F}_{\rm M} , \\  
\label{ceffeq}
\ceffsq & = & \sign^2 \frac{(3 \Pexttil + \sign^2)}{(\Pexttil + \sign^2)^2} , \\
\label{rhon}
\rhon & = & \frac{1}{4} \left( \sign^2 + \Pexttil + \bl{B}_p^2 \right),\\
\label{Zeq}
Z & = & \frac{\sign}{2 \rhon}, \\                                
\bl{g}_p & = & -\nabla_p \psi  ,\\
\label{gravpot}
\psi & = & \FTinv \left[ - \FT(\sign)/k_z \right]     ,\\
\bl{B}_p & = & -\nabla_p \Psi  ,\\
\label{magpot}
\Psi & = & \FTinv \left[ \FT(\Beq - \Bref)/k_z \right]  \, .
\earray
In the above equations, $\sign(x,y) = \int_{-Z}^{+Z}\rhon(x,y)~dz$ is 
the column density of neutrals, 
$\bl{B}_p(x,y) = B_x(x,y)\hat{\bl{x}} + B_y(x,y)\hat{\bl{y}}$ is the 
planar magnetic field at the top surface of the sheet, 
$\vnp(x,y) = v_x(x,y)\hat{\bl{x}} + v_y(x,y)\hat{\bl{y}}$ is the 
velocity of the neutrals in the plane, 
$\vip(x,y) = v_{{\rm i},x}(x,y)\hat{\bl{x}} + v_{{\rm i},y}(x,y)\hat{\bl{y}}$ 
is the corresponding velocity of the ions, and 
the normalized initial mass density (in units of $\signi/L_0$)
$\rhoni = \frac{1}{4}(1+\Pexttil)$, where $\Pexttil$ is defined below. 
The operator $\nabla_p = \hat{\bl{x}} \, \partial/\partial x + 
\hat{\bl{y}} \, \partial/\partial y$ is the gradient in the planar
directions within the sheet.
The quantities $\psi(x,y)$ and $\Psi(x,y)$ are the 
scalar gravitational and magnetic potentials, respectively, also in the plane
of the sheet. The vertical wavenumber $k_z = (k_x^2+k_y^2)^{1/2}$ is 
a function of wavenumbers $k_x$ and $k_y$ in the plane of the sheet,
and the operators $\FT$ and $\FTinv$ represent the forward
and inverse Fourier transforms, respectively, which we calculate 
numerically using an FFT technique. Terms of order 
${\cal O}(\nabla_p Z)$ in $\bl{F}_{\rm M}$, the magnetic force per
unit area, are not 
written down for the sake of brevity, but are included in the numerical
code; their exact form is given in Sections 2.2 and 2.3 of CB06. All terms 
proportional to $\nabla_p Z$ are generally very small.

We also note that the effect of nonzero $\nabla_p Z$ and
external pressure $\Pext$ is accounted for in the vertically-integrated 
thermal pressure force per unit area, $\bl{F}_{\rm T}$, through the use of 
$\ceffsq$. This can be seen by noting that
\barray
\label{intP}
{\bl F}_{\rm T} = \int_{-Z}^{+Z} \nabla_p P~dz 
&=& \nabla_p \int_{-Z}^{+Z} P~dz - 2 \Pext \nabla_p Z \nonumber \\ 
&=& 2 \, \nabla_p (P Z - \Pext Z)~~,
\earray
where $P$ is the pressure inside the sheet. In the above expression, 
we have used $P=\Pext$ at the upper and lower surfaces of the sheet, 
and also that $\nabla_p (+Z) = -\nabla_p (-Z)$. 
Using the ideal gas equation for an isothermal gas, $P=\rhon\,\csq$, the 
expression for half-thickness (Eq. [\ref{Zeq}]), and the normalized
equation for vertical
hydrostatic equilibrium (Eq. [\ref{rhon}], where we ignore the relatively small
term $\bl{B}_p^2$ for simplicity), it is straightforward to 
derive the normalized expression for $\ceffsq$ (Eq. [\ref{ceffeq}]).

The above equations contain the following dimensionless free parameters:
$\Pexttil \equiv 2 \Pext/\pi G\signi^{2}$ is the ratio of the external 
pressure acting on the sheet to the vertical self-gravitational stress 
of the reference state. The dimensionless neutral-ion collision time
of the reference state, $\tniitil \equiv \tnii/t_0$, 
expresses the effect of ambipolar diffusion. 
In the limit $\tniitil \rightarrow \infty$ there
is extremely poor neutral-ion collisional coupling, such that the ions 
and magnetic field have no effect on the neutrals. The opposite limit,
$\tniitil =0$, corresponds to the neutrals being perfectly coupled
to the ions due to frequent collisions, i.e. flux freezing.
The neutral-ion collision time of the reference state is
\beq
\tnii = 1.4 \frac{\mi + m_{{}_{\Htwo}}}{\mi} \frac{1}{\nii \sigin}\;,
\label{tni}
\eeq
where $\mi$ is the ion mass, which we take to be 25 a.m.u.,
the mass of the typical atomic ($\rm{Na}^{+}$, $\rm{Mg}^{+}$)
and molecular ($\rm{HCO}^{+}$) ion species in clouds, 
$\nii$ is the ion number density of the reference state,
and $\sigin$ is the neutral-ion collision rate, equal to
$1.69 \times 10^{-9}~{\rm{cm}}^{3}~{\rm s}^{-1}$ for 
$\Htwo$-${\rm{HCO}}^{+}$ collisions \citep{McDa}. 
The factor of 1.4 in Eq. (\ref{tni}) accounts for the fact
that the effect of helium is neglected in calculating the slowing-down
time of the neutrals by collisions with ions.
The parameter $\kion$ is the exponent in the power-law expression that is used 
to calculate the ion density $\nion$ as a function of neutral density
$\nn$, namely,
\beq
\label{rhoieq}
\nion = {\cal K} \nn^{\kion}~,
\eeq
where we adopt 
$\kion = 1/2$ and ${\cal K}\, (\simeq 10^{-5} {\rm cm}^{-3/2})$ for all models 
in this study \citep[e.g.,][]{Elme79, UN80}, but keep in mind that 
calculation of the ion chemistry network makes $\kion$ a function of $\nn$
\citep{CM98}. Finally, $\Breftil = \Bref/B_0 = \Bref/2 \pi G^{1/2} \signi$ is the 
dimensionless magnetic field strength of the reference state. For physical
clarity, we use instead the
dimensionless mass-to-flux ratio of the background 
reference state:
\beq
\label{muieq}
\mui \equiv 2 \pi G^{1/2} \frac{\signi}{\Bref} = 
\Breftil^{-1} \; ,
\eeq
where $(2 \pi G^{1/2})^{-1}$ is the critical mass-to-flux ratio for 
gravitational collapse in our adopted thin-sheet geometry (CB06). Models with 
$\mui < 1$ ($\Breftil > 1$) are subcritical clouds, 
and those with $\mui > 1$ ($\Breftil < 1$) are supercritical.
The initial mass-to-flux ratio is also related to the commonly-used plasma 
parameter
\beq
\beta_0 \equiv \frac{\rhoni\, \csq}{(\Bref^2/8\pi)} = \mui^2 \, (1+ \Pexttil).
\eeq 


Typical values of our units are
\barray
\label{cs}
\cs & = & 0.188 \, \left(\frac{T}{10 \K}\right)^{1/2} \kms,  \\
\label{t0}
t_0 & = & 3.65 \times 10^4 \left(\frac{T}{10\,\K}\right)^{1/2}
\left(\frac{10^{22}\,\cms}{\Nni}\right) \yr, \\
\label{L0}
L_0 & = & 7.02 \times 10^{-3} \left(\frac{T}{10 \K}\right)
\left(\frac{10^{22}\,\cms}{\Nni}\right) \pc \\ \nonumber
& = & 1.45 \times 10^3  \left(\frac{T}{10 \K}\right)
\left(\frac{10^{22}\,\cms}{\Nni}\right) \AU, \\
\label{M0}
M_0 & = & 9.19 \times 10^{-3} \left(\frac{T}{10\,\K}\right)^2
\left(\frac{10^{22}\,\cms}{\Nni}\right) \Msun, \\
\label{B0}
B_0 & = & 63.1 \left(\frac{\Nni}{10^{22}\,\cms}\right) \muG.
\earray
Here, we have used $\Nni = \signi/\mn$, where $\mn = 2.33\,\mH$ is the mean
molecular mass of a neutral particle for an H$_2$ gas with a 10\% 
He abundance by number. 
Furthermore, we may calculate the number density of the background state as
\beq
\nni = 2.31 \times 10^5 
\left(\frac{10 \K}{T}\right) 
\left(\frac{\Nni}{10^{22}\cms}\right)^2
\left(1+\Pexttil\right)\, \cmc.
\eeq
The dimensional background reference magnetic field strength for a given model is simply 
$\Bref = B_0/\mui$.
Finally, the ionization fraction ($=\nion/\nn$) in the cloud 
may be expressed as
\beq
\label{ioniz}
\xion = {\cal K} \nn^{-1/2} =  3.45 \times 10^{-8} \left(\frac{0.2}{\tniitil}\right) \left(\frac{10^5 \cmc}{\nn}\right)^{1/2} \left(1+\Pexttil\right)^{-1/2}.
\eeq

\subsection{Numerical Techniques, Boundary and Initial Conditions}
\label{s:techn}

The system of Eqs. (\ref{cont}) - (\ref{magpot}) 
are solved numerically in $(x,y)$ coordinates using a multifluid 
non-ideal MHD code that was 
specifically developed for this purpose (BC04; CB06). 
Partial derivatives $\partial/\partial x$ and $\partial/\partial y$ are
replaced with their finite-difference equivalents. 
Gradients are approximated using three-point central 
differences between mesh cells, while advection of mass and magnetic 
flux is prescribed by using the monotonic upwind scheme of \citet{vanL}.
Evolution of a model is carried out 
within a square computational domain of size $L \times L$, spanning the
region $-L/2 \leq x \leq L/2$ and $-L/2 \leq y \leq L/2$. Typically, $L$
is taken to be several times larger (up to a factor of 4) than the characteristic
length scale of maximum gravitational instability $\lamgm$ 
(CB06; see, also, Section~\ref{s:results} below). The computational domain is then
divided into a set of $N^2$ equally-sized mesh cells, each having an 
area $L/N \times L/N$. 
Most of our simulations are run with $L=16\pi\,L_0$ and $N=128$, and some
have $L=64\pi\,L_0$ and $N=512$, so that the grid size $\Delta x =
\Delta y = 0.393\,L_0$ in all cases. The mass resolution is then
$\Delta M = 0.154\,M_0$, or $1.42 \times 10^{-3}\,\Msun$ using the standard values in
Eq. (\ref{M0}).

The numerical 
method of lines \citep{Schi} is employed, i.e. the first-order
partial differential equations
(\ref{cont}) - (\ref{induct}) are converted into a set of coupled 
ordinary differential equations (ODE's) in time, with one ODE for each
physical variable at each cell. Hence, the system of ODE's 
has the form $d\bl{\cal Y}/dt = \bl{\cal G}(\bl{\cal Y},t)$,
where $\bl{\cal Y}$ and $\bl{\cal G}$ are both arrays of size $V N^2$, 
$V$ being the number of dependent variables. Time-integration of this 
system of ODE's is performed by using an Adams-Bashforth-Moulton 
predictor-corrector subroutine \citep{Sham94}. Numerical solution of 
Fourier transforms and inverse transforms, necessary to calculate the 
gravitational and magnetic potentials $\psi$ and $\Psi$ at each time 
step (see Eqs. [\ref{gravpot}] and [\ref{magpot}]), is done by using 
fast Fourier transform techniques \citep{Press}. 

Periodic conditions are applied to all physical
variables at the boundary of the computational domain.
The background reference state of a model cloud is characterized
by a uniform column density
$\signi$ and magnetic field $\Beqi \zhat = \Bref \zhat$. This means that
the gravitational and magnetic forces are each identically
zero in the uniform background state. The evolution of a model cloud 
is started at time $t=0$ by superposing a set of perturbations 
$\delta\sign(x,y)$ that are random white noise with a root-mean-squared (rms)
value that is 3\% of $\signi$. To preserve the
same local mass-to-flux ratio $\sign/\Beq$ as in the uniform background
state, initial magnetic field perturbations 
$\delta\Beq = (\delta\sign/\signi)\Bref$ are also introduced.

Detailed tests of the accuracy of this MHD code were described in CB06. 
Full code runs were compared to exact linear solutions for the
gravitationally unstable modes of thin-sheet magnetic clouds. The 
code was found to be in excellent agreement with these solutions.
It correctly captured the temporal evolution of a model cloud in
the linear regime of collapse, as exemplified by the growth time of the
gravitational instability $\tau_{\rm g}$ for a given fragmentation 
length scale $\lambda$, for various values of the
initial parameters $\mui$, $\tniitil$, and $\Pexttil$.
In addition, we have run flux-freezing tests of subcritical models
with random perturbations and verified that no spurious gravitational
instability occurs in the absence of ambipolar diffusion.

The principal motivation for our modeling clouds as thin sheets is that
it significantly reduces the computational complexity of studying star 
formation, while still retaining many fundamental physical features 
necessary to understanding the dynamics of core formation and collapse 
within interstellar clouds. For instance, although model clouds are 
thin, they are not infinitesimally so, and we are able to 
incorporate both magnetic pressure and magnetic tension 
supporting forces (see Eq. [\ref{Fmag}]).
Additionally, the vertical integration (along the direction of the 
$z$-axis) that is employed to derive the system of governing equations 
in the thin-sheet approximation (Eqs. [\ref{cont}] - [\ref{magpot}]) 
has the effect of turning the fully three-dimensional gravitational 
collapse problem into a computationally more tractable 
two-dimensional problem. 
As a result of these computational
savings, our numerical code is able to run efficiently on 
a single workstation with minimal cpu times. 
Depending on the initial parameters, a 
full simulation can be completed in as little
as an hour or at most a single day. Hence, we are able to quickly 
generate a large number of models covering the entire physically 
relevant range of our free parameters, and
produce a large quantity of models that can be used for statistical 
analysis. By contrast, three-dimensional MHD models
require a dedicated workstation or a computer cluster, and 
their simulation completion times are orders of magnitude greater than 
that needed for our thin-sheet models. 
Our new code is written in the IDL programming language,
which significantly speeds up the processes of code development,
debugging, and visualization.

We have also developed software to analyze the masses and shapes of cores
arising from our simulations. In any snapshot of the evolution, we first
isolate regions with column density (or mass-to-flux ratio in some cases)
above some threshold value. Thresholds are usually chosen to be high
enough that multiple peaks do not fall within a single contiguous
region above the threshold. In cases where this happens,
we have the option of manually isolating the cores.
The mass of each core is found by adding the masses of each 
computational zone in the isolated region above the threshold. 
To determine the size and shape of a core we use the MPFITELLIPSE
routine written in IDL by C. Markwardt, which returns the best-fit ellipse
to the set of zones that constitute each core. The semimajor
and semiminor axes of the best-fit ellipse, $a$ and $b$, respectively, are obtained
and used to determine the size $s = \sqrt{ab}$ and axis ratio $b/a$
of each core. The vertical half-thicknesses $Z$ of the zones
within each core are averaged to find a mean half-thickness.
The mean separation of fragments are found by locating,
for each density peak associated with a core, the nearest peak
of a neighboring core. These values are averaged over all cores in a simulation 
and over multiple model realizations. The periodic boundary 
conditions are also accounted for; we count any possible
nearest neighbor that is just across the periodic boundary.

\section{Results}
\label{s:results}

\subsection{Overview}
\label{s:over}

The efficiency of our two-dimensional code allows us to run a large number
of simulations, with various combinations of the important parameters
$\mui$, $\tniitil$, and $\Pexttil$. For each unique set of parameters
we are also able to run a multitude of independent model 
realizations. Each realization is distinct in specific details,
since the evolution is initiated by random (white noise) small-amplitude 
perturbations. However, the independent realizations are 
statistically similar, and running a large number of models allows
us to assess the level of randomness that contributes to 
distributions of various calculated quantities. 

Table 1 contains the parameters for each of ten models as well as
the predicted minimum growth time $\taugm$, the associated wavelength of 
maximum instability $\lamgm$, and the implied fragmentation mass
$\mgm \equiv \pi \signi \lamgm^2/4$, all obtained 
from the linear perturbation analysis of CB06.
The first two quantities are in normalized form, but the masses have
been converted to $\Msun$ using the standard values in Eq. (\ref{M0}).
Some insight into the numerical values of $\taugm$ and $\lamgm$ in Table 1
can be obtained by considering the growth rate and fragmentation
scale of the fastest growing mode in the limit of no magnetic
field, but finite external pressure. From the results of CB06, we 
find
\barray
\label{taug}
\tau_{\rm g}(\Pexttil,\Breftil=0) & = & 2 \,\frac{(1+3\Pexttil)^{1/2}}{(1+\Pexttil)} \frac{L_0}{\cs} = (1+3\Pexttil)^{1/2} \,\frac{Z_0}{\cs}, \\
\label{lamg}
\lambda_{\rm g}(\Pexttil,\Breftil=0) & = & 4 \pi \frac{(1+3\Pexttil)}{(1+\Pexttil)^2} L_0
= 2 \pi \left( \frac{1+3\Pexttil}{1+\Pexttil}\right) Z_0. 
\earray
In the above equations, we have used the relation
\beq
Z_0 = \frac{2L_0}{(1+\Pexttil)}.
\eeq
These results show that in the limit $\Pexttil \rightarrow 0$, the isothermal
sheet has effective ``Jeans length'' $\lammax = 4 \pi L_0 = 2 \pi Z_0$,
and growth time $\tau_{\rm T,m} = 2L_0/\cs = Z_0/\cs$. The
relation of these values to the sheet half-thickness
and sound speed is intuitively understandable; the latter is essentially
the dynamical time. 
The highly supercritical 
model 10, which also has low external pressure, does approach these limiting 
values of fragmentation scale and growth time.
Equations (\ref{taug})-(\ref{lamg}) also bring out the interesting property that the fragmentation
length and time scale of the isothermal sheet can be reduced significantly by
increases in $\Pexttil$, but that the fragmentation scale converges to 
a fixed multiple of $Z_0$ (i.e. $6 \pi Z_0$) as $\Pexttil \rightarrow \infty$.
This property was first noted by \citet{Elme78} and studied further by \citet{Lubo}.
The significantly subcritical model 1 also has a fragmentation
scale $\approx 2 \pi Z_0$, since instability occurs via neutral drift 
past near-stationary field lines; however the growth time $\taugm \approx 10\,Z_0/\cs$. 
The transcritical models 3 and 4
have values of $\taugm$ more similar to the subcritical models than the
supercritical ones, although their fragmentation scales $\lamgm$ are large.
Models 9 and 10 have have $\Pexttil=10$,
resulting in much smaller values of $\taugm$ and $\lamgm$ than corresponding
models with $\Pexttil=0.1$, although the dynamically important magnetic field
raises the values of both above the nonmagnetic limits.

Our standard simulation box is four times larger in size than
$\lammax$ and more than twice $\lamgm$ for most models.
Exceptions to this are model 3 and model 7, and these simulations are
carried out in larger simulation boxes of quadruple size ($L=64\,\pi L_0,N=512$)
when collecting information on core properties.
The total mass in the simulation box with 
$L=16\,\pi L_0 = 0.353 \,( \Nni /10^{22}\cms)^{-1} (T/10 \K)$ pc is 
$M = 2.53 \times 10^3\, M_0 
= 23.2\, ( \Nni /10^{22}\cms)^{-1}(T/10 \K)^2 \,\Msun$.
The expected fragmentation scales and masses for all models are 
significantly greater than the grid length resolution $\Delta x$ and
mass resolution $\Delta M$ quoted in Section \ref{s:techn}.



Table 2 contains model parameters and key quantities at
the end of each nonlinear model run.
All runs end when $\signmax/\signi = 10$. This corresponds to 
a volume density enhancement $\rhon/\rhoni \approx 100$ for models with
$\Pexttil=0.1$ and $\rhon/\rhoni \approx 10$ for models with $\Pexttil=10$.
For each model, we list representative values of $t_{\rm run}$\footnote{Experimentation
with different rms amplitudes of the initial perturbation (so that 
$\delta \sign/\signi$ is in the range 1\%-6\%), and variation of the power
spectrum away from white noise but with the fixed standard rms value, reveal
that the values of $t_{\rm run}$ can vary in the range 10\%-20\% from
the values quoted in Table 2.},
the physical time at which $\signmax=10\signi$,
$\vnmax$, the maximum neutral speed in the simulated region at that time,
and $\vimax$, the corresponding quantity for the ions. 
In general, models with relatively large values of 
$\mui, \tniitil, ~{\rm and}~ \Pexttil$ tend
to evolve faster than counterparts with smaller values of these parameters.
For the gravity-dominated models ($\Pexttil=0.1$), increasing values of $\mui$
and/or $\tniitil$ result in greater values of $\vnmax$.
The values of $\vimax$ illustrate that the systematic motions
within the nonlinearly developed cores are gravitationally driven, 
so that ions lag behind neutrals somewhat. 
However, the difference between the speeds of the two species 
remains less than $0.1\,\cs$.
We perform an analysis of core properties as described in Section \ref{s:techn},
after defining a core as an enclosed region with $\sign/\signi \geq 2$
at the end of the simulation.
Since each simulation typically yields only a handful of cores, the
models are run a large number of times to generate significant
core statistics. Models 1, 2, 3, and 5 were run 100 times each, while models
4 and 7, which need to be run on an expanded grid due to very large
fragmentation scales, were run 22 and 6 times, respectively.
Model 8 ran 50 times and models 9 and 10 were run 25 times each.
Several averaged properties of the resulting cores are presented in columns
8-12 of Table 2. These quantities are 
$\lamavg$, the average distance between cores,
$\Mavg$, the average mass within a core (converted to $\Msun$
using Eq. [\ref{M0}]), $\savg$, the average size
of a core, $\axisravg$, the average 
axis ratio of a core, and $\Zavg$, the average value of the half-thickness 
of a core. 





\begin{table}
\begin{center}
\caption{Summary of Parameters and Results of Linear Theory}
\end{center}
\label{t:1}
\begin{tabular}{crrrrrr}
\hline \hline
Model &\hspace{1em}$\mu_0$ &$\hspace{1em}\tniitil$ &\hspace{1em}$\Pexttil$ &\hspace{2em}$\taugm$ &\hspace{1em}$\lamgm$ & $\mgm$ 
\\
\hline
1       &0.5     &0.2    &0.1    &20.2  &14.3 &1.48
\\
2       &0.8     &0.2    &0.1    &17.6  &16.5 &1.96
\\
3       &1.0     &0.2    &0.1    &14.3  &24.3 &4.26
\\
4       &1.1     &0.2    &0.1    &10.8   &54.9 &21.8
\\
5       &2.0     &0.2    &0.1    &3.21   &23.9 &4.12
\\
6       &10.0    &0.2    &0.1    &2.11   &13.9  &1.40
\\
7       &1.0     &0.1    &0.1    &27.5   &28.5  &5.86
\\
8       &1.0     &0.4    &0.1    &7.93   &20.0 &2.89
\\
9       &0.5     &0.2    &10.0   &4.87   &3.43 &0.09
\\
10      &1.0     &0.2    &10.0   &3.33   &4.99 &0.18

\\
\hline
\end{tabular}
\\
Times and lengths are normalized to $t_0$ and
$L_0$, respectively. Masses are converted to $\Msun$ using Eq. (\ref{M0}).
\end{table}

\begin{table}
\begin{center}
\caption{Main Results for Model Clouds}
\end{center}
\label{t:2}
\begin{tabular}{crrrrrrrcrrr}
\hline \hline
Model &\hspace{1em}$\mu_0$ &$\hspace{1em}\tniitil$ &\hspace{1em}$\Pexttil$ &\hspace{2em}$t_{\rm run}$&\hspace{1em}$\vnmax$& $\hspace{1em}\vimax$ &\hspace{3em}$\lamavg$ &\hspace{1.5em}$\Mavg$ &\hspace{1.5em}$\savg$ &\hspace{1.5em}$\axisravg$ &\hspace{1.5em}$\Zavg$ 
\\
\hline
1       &0.5     &0.2    &0.1    &204  &0.39  & 0.35 &15.0  &\hspace{1.5em}1.26  &1.25  &0.74 
&0.71\\
2       &0.8     &0.2    &0.1    &167  &0.62  &0.55 &17.7  &\hspace{1.5em}1.56  &1.49  &0.81 
&0.72\\
3       &1.0     &0.2    &0.1    &121  &0.70   & 0.64 &20.1  &\hspace{1.5em}3.26  &2.09  &0.69
&0.65\\
4       &1.1     &0.2    &0.1   &88   &0.95  &0.90 &47.1  &\hspace{1.5em}6.62
&3.23  &0.66  &0.75\\
5       &2.0     &0.2    &0.1    &23   &1.1  &1.0 &19.1  &\hspace{1.5em}3.41  &2.08  &0.57
&0.67\\
6       &10.0    &0.2    &0.1    &12   &1.2  &1.2 &12.8  &\hspace{1.5em}0.99  &1.04  &0.53
&0.70\\
7       &1.0     &0.1    &0.1   &261 &0.66  &0.62 &31.4  &\hspace{1.5em}3.29
&2.13  &0.77  &0.78\\
8       &1.0     &0.4    &0.1    &61  &0.73  &0.63 &18.5  &\hspace{1.5em}2.02  &1.69  &0.67
&0.71\\
9       &0.5     &0.2    &10.0   &44   &0.70  &0.63 &4.0   &\hspace{1.5em}0.34  &0.60  &0.62
&0.072\\
10      &1.0     &0.2    &10.0   &22   &0.50  &0.38 &5.3   &\hspace{1.5em}0.46  &0.74  &0.60
&0.074\\
\hline
\end{tabular}
Times and lengths are normalized to $t_0$ and
$L_0$, respectively. Speeds are normalized to $\cs$. 
Masses are converted to $\Msun$ using Eq. (\ref{M0}).
Core data for models 4 and 7 are compiled from runs with $N = 512, L = 64\pi L_0$.
\end{table}

\bfig
\epsfig{file=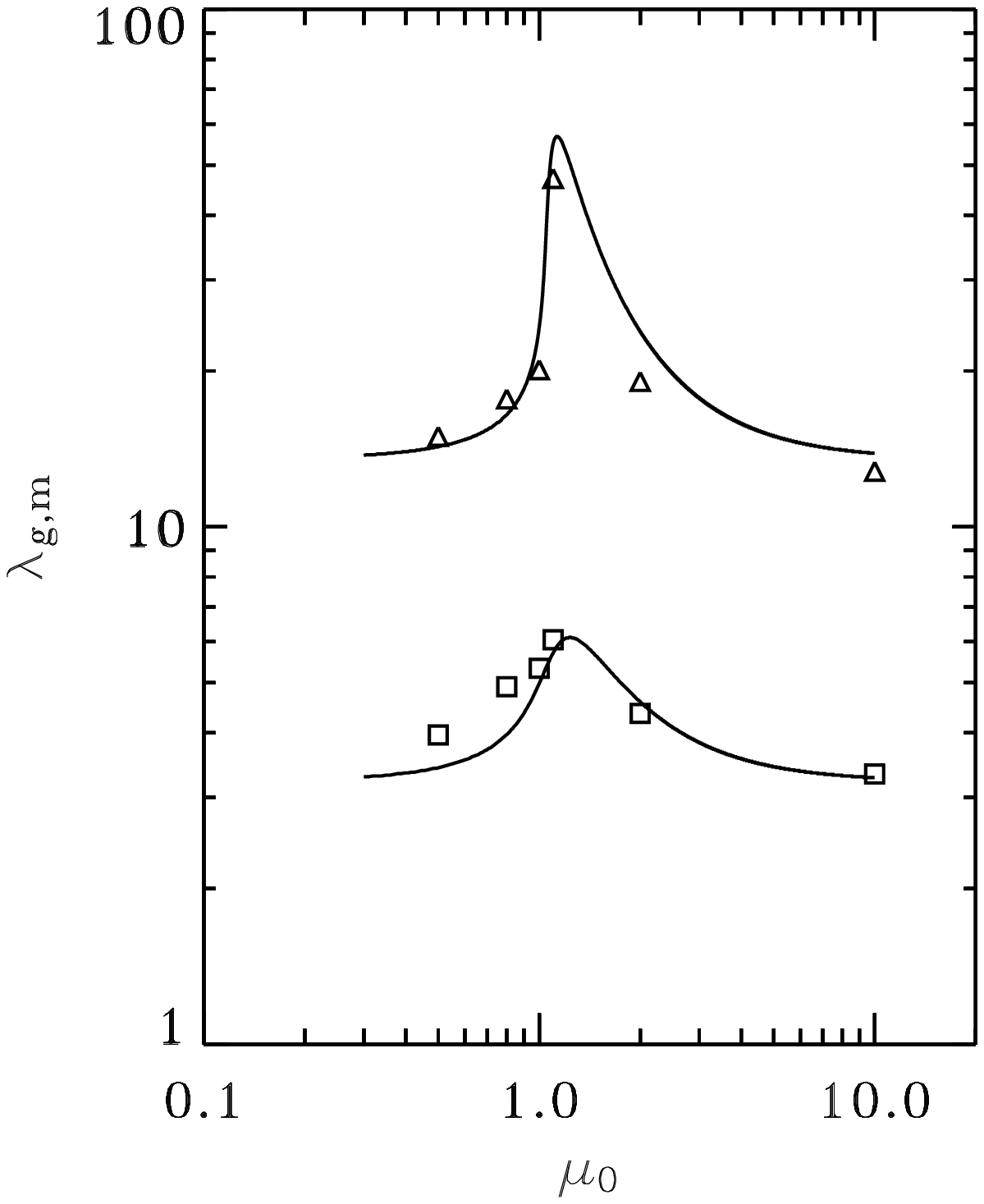}
\caption{Preferred fragmentation scales $\lamgm$ from linear stability 
analysis \citep{CB06} compared with
the average spacing of density peaks in the nonlinear simulations.
The upper solid line is the calculated dependence of
the wavelength with maximum growth rate $\lamgm$ versus 
$\mui$ for fixed parameters $\tniitil=0.2$ and $\Pexttil=0.1$.
The lower solid line is the same but for $\Pexttil = 10$. The triangles 
represent the average spacing of density peaks (with peak $\sign \geq 2 
\signi$) tabulated from a large number
of simulations at each of $\mui = 0.5,0.8,1.0,1.1,2.0$, and $10.0$ with
$\tniitil=0.2$ and $\Pexttil=0.1$. The squares represent the same but 
with $\Pexttil = 10$.}
\label{scales}
\efig

Fig.~\ref{scales} shows the preferred fragmentation scales $\lamgm$ versus $\mui$ 
from the linear analysis of CB06 for two separate values 
of the dimensionless external pressure ($\Pexttil = 0.1$ on top 
and $\Pexttil = 10$ below). Both lines represent models with
$\tniitil = 0.2$, which is our adopted standard value based on 
typical observationally-inferred ionization levels (see Eq. [\ref{ioniz}]
above and Eq. [29] of CB06).
Overlaid on each solid line are 
average core spacings $\lamavg$ (see Table 2) calculated from  
the nonlinear endpoint of our simulations. Each data point 
represents an average of core spacings for a large number (20 to 100) of 
simulations. 
The lower line contains $\lamavg$ data for up to 25 runs each of 
an additional four models
with $\Pexttil=10$ that do not, for the sake of brevity, have their data 
compiled in Tables 1 and 2.
{\it These results show that the linear theory can be used with
confidence to predict the average fragmentation properties of clouds even
in a fully nonlinear stage of development.} 
The tabulated fragmentation scales are slightly below the predictions
of linear theory in the range $\mui \approx 1-2$, for $\Pexttil=0.1$.
This is due to occasional subfragmentation of the initially large 
(irregularly shaped) fragments as they become decidedly supercritical.
We return to this issue when discussing the models with
$\mui=1.1$.


\subsection{The Effect of Varying $\mui$} 
\label{s:mui}

\subsubsection{Time Evolution}
\label{s:timevol}

Fig.~\ref{timevol} shows the time evolution of the maximum neutral column 
density $\signmax$ and maximum mass-to-flux ratio $\mu_{\rm max}$,
both normalized to their initial values, for models 1, 3, and 5, 
which have $\mui= 0.5,1.0,$ and 2.0, respectively, and fixed parameters
$\tniitil=0.2$ and $\Pexttil=0.1$. These three models have
$\taugm/t_0=20.2,14.3,$ and 3.2, respectively. The actual time $t_{\rm run}$
to runaway collapse of a core, starting from small-amplitude
white-noise perturbations, is about $7-10$ times $\taugm$ for these
models. Review of Table 1 and Table 2 shows that $t_{\rm run}/\taugm
\approx 7-10$ is a generic feature of all our models. The clouds with
initial critical and subcritical mass-to-flux ratio have a prolonged
period of dormancy, compared to the supercritical models.
This is due to the need for ambipolar diffusion to operate before
collapse sets in for both cases. The dashed lines in Fig.~\ref{timevol} show how 
much the mass-to-flux ratio changes during the evolution.
The initially subcritical
cloud requires a significant increase of $\mu_{\rm max}$
before collapse begins. 
Although $\mu_{\rm max}$ appears to be diverging
at the end of the simulations, it is in fact increasing much more 
slowly than $\signmax$, and will not asymptotically diverge. This can
be seen in previously published work, i.e. in Fig. 2 of each of
\citet{CM94} and \citet{BM94}.

\bfig
\epsfig{file=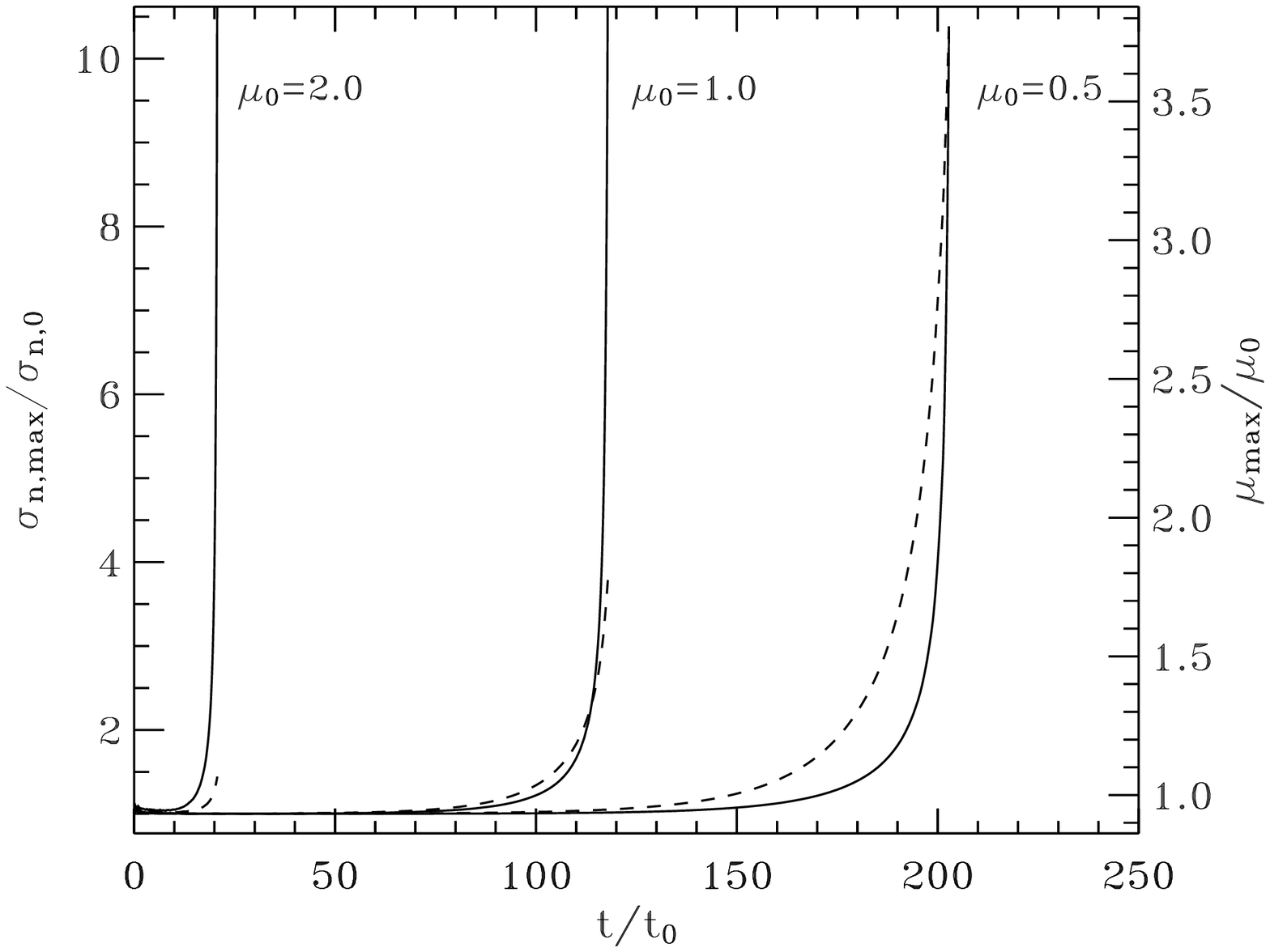,width=\linewidth}
\caption{Time evolution of maximum values of surface density and
mass-to-flux ratio in three simulations. The solid lines show the
evolution of the maximum value of surface density in the simulation,
$\signmax/\signi$, versus time $t/t_0$. This is shown for models
1, 3, and 5, which have initial mass-to-flux ratio values
$\mui=$ 0.5, 1, and 2, respectively. For each model, a dashed line
shows the evolution of the maximum mass-to-flux ratio in the simulation,
$\mu_{\rm max}$, normalized to $\mui$.
}
\label{timevol}
\efig

\subsubsection{Column Density and Velocity Structure}
\label{s:cdens}

Fig.~\ref{densimgs} shows the column density map and velocity vectors 
of neutrals at the
end of the simulations for parameters $\mui=0.5,0.8,1.0,1.1,
2.0,~{\rm and}~10.0$, with $\tniitil=0.2$ and $\Pexttil=0.1$.
All simulations end when $\signmax/\signi=10$, but the time at which 
this is reached is different in each model. These times for the various
models (in order of increasing $\mui$) are $t/t_0=203.7,166.5,120.9,88.1,22.8,~{\rm and}~12.4$.
There is a striking variation in the spacings of cores as $\mui$ 
changes. The highly subcritical case $\mui=0.5$ fragments on essentially
the nonmagnetic preferred scale $\lammax = 4 \pi L_0$, since the 
evolution is characterized by diffusion of neutrals past near-stationary
magnetic field lines, i.e. a Jeans-like instability but on a diffusive time scale. 
As predicted by the linear theory (CB06), there is a peak in the 
fragmentation spacing near $\mui=1$. The peak 
occurs at $\mui=1.1$ when $\tniitil=0.2$ and $\Pexttil=0.1$.
The predicted fragmentation scale is $\lamgm=4.2\,\lammax$,
and indeed our simulation with box width $4\,\lammax$ yields only
one core, although it seems to be undergoing a secondary
fragmentation into two pieces.

The velocity vectors of the neutral flow are normalized to the same
scale in each frame, and the horizontal or vertical spacing of the
footpoints is equal to $0.5\,\cs$. There is a monotonic increase of
the typical neutral speeds as $\mui$ increases (see values of $\vnmax$
in Table 2). The supercritical models, unlike their subcritical and
critical counterparts, have large-scale flow patterns
with velocities in the approximate range $(0.5-1.0)\,\cs$, and 
maximum speeds associated with the most fully developed cores that
are mildly supersonic at distances $\sim 0.1$ pc from the core centers.
Interestingly, the essentially hydrodynamic model with $\mui=10$
has only somewhat greater systematic speeds than the model with $\mui=2$.
This is because the fragmentation scale is smaller, so that 
each core has a weaker gravitational influence on its surroundings at this stage
of development.
However, recall that the frames are at different physical times 
since the models with smaller values of $\mui$ reach $\signmax/\signi=10$
at progressively later times.
Spatial profiles of $\vnp$ in the vicinity of cores are presented in BC04, for
some models, and we do not present them again in this paper. The trend of values
of maximum neutral speed $\vnmax$ and maximum ion speed $\vimax$ (Table 2) reveal the 
predictions of our models for the observable motions on the core scale. 
The infall motions are gravitationally driven, so that the ion speed lags
the neutral speed in all cases. The overall rms speed in the entire simulation
region is always quite small for both neutrals and ions. The rms speed of
neutrals, $v_{\rm n,rms}$, falls in the range $(0.05-0.21)\,\cs$ for the various
models, and the corresponding quantity $v_{\rm i,rms}$ in the range
$(0.04-0.20)\,\cs$.



Since the standard box size $L =16\, \pi L_0$ allows only one fragment to 
form initially in the model with $\mui=1.1$ (as seen in Fig. \ref{densimgs}), we ran
another model with four times larger box size but the same resolution, 
so that $L=64\, \pi L_0$ and $N=512$. This allows the formation of multiple
fragments, since the preferred fragmentation scale from the linear theory
is $\lamgm=54.9 L_0$. Fig.~\ref{bigimage} shows the column density and velocity vectors
at the end of one such simulation. Velocity vectors are again normalized such 
that the horizontal or vertical spacing between footpoints equals $0.5\,\cs$.
This larger simulation does show that multiple fragments form with spacings
approximately as predicted by the linear theory, but that there is also
a tendency for cores to subfragment into two density peaks.
An analysis of the result of 25 separate simulations with different 
random realizations of the initial state reveals that the average distance
between density peaks is $47.1 L_0$, which is somewhat smaller than $\lamgm=54.9 L_0$.
This is explained by the occasional presence of secondary density peaks
within the initially formed fragments. 
While most model clouds undergo {\it single-stage fragmentation} into essentially
thermal critical (Jeans-like) fragments of size $\approx \lammax$, the 
transcritical clouds form
first-stage fragments many times larger than $\lammax$, followed by a possible
{\it second-stage fragmentation} when the mass-to-flux ratio of the fragment 
becomes decidedly supercritical due to ambipolar diffusion. 
This second-stage fragmentation may be favored because both the preferred
fragmentation
scale and growth time of gravitational instability drop precipitously as a cloud
makes the transition from transcritical to supercritical (see Figs. 1{\it d} and 2 
of CB06). The initially rather large fragment may itself
be prone to fragmentation because of its irregular shape.


Fig.~\ref{surfaceplots} shows an alternate view of the column density, using surface plots
at the end of simulation
runs, with parameters corresponding to those of models 1, 3 and 5. 
These models start from different realizations of the initial state
than those of the models presented in Fig.~\ref{densimgs}. Density peaks occur in 
different locations but represent an equivalent outcome statistically.
Animations of the time evolution of the surface plots are 
available online\footnote{The animations of the models shown in Figs.~\ref{surfaceplots} and \ref{fls}
reveal that they reach the final state with $\sign/\signi=10$
in a time 15-20\% less than that quoted in Table 2. This is because
the initial perturbation is white noise but with the smallest wavelengths
($\lambda \leq 4$ grid cells) damped out. This results in slightly more power
in the longer wavelengths (for a fixed rms perturbation level), 
including the preferred fragmentation scale $\lamgm$, and a consequent 
quicker development of the favored mode.}.

\bfig
\vspace{-15ex}
\centering
\begin{tabular}{cc}
\epsfig{file=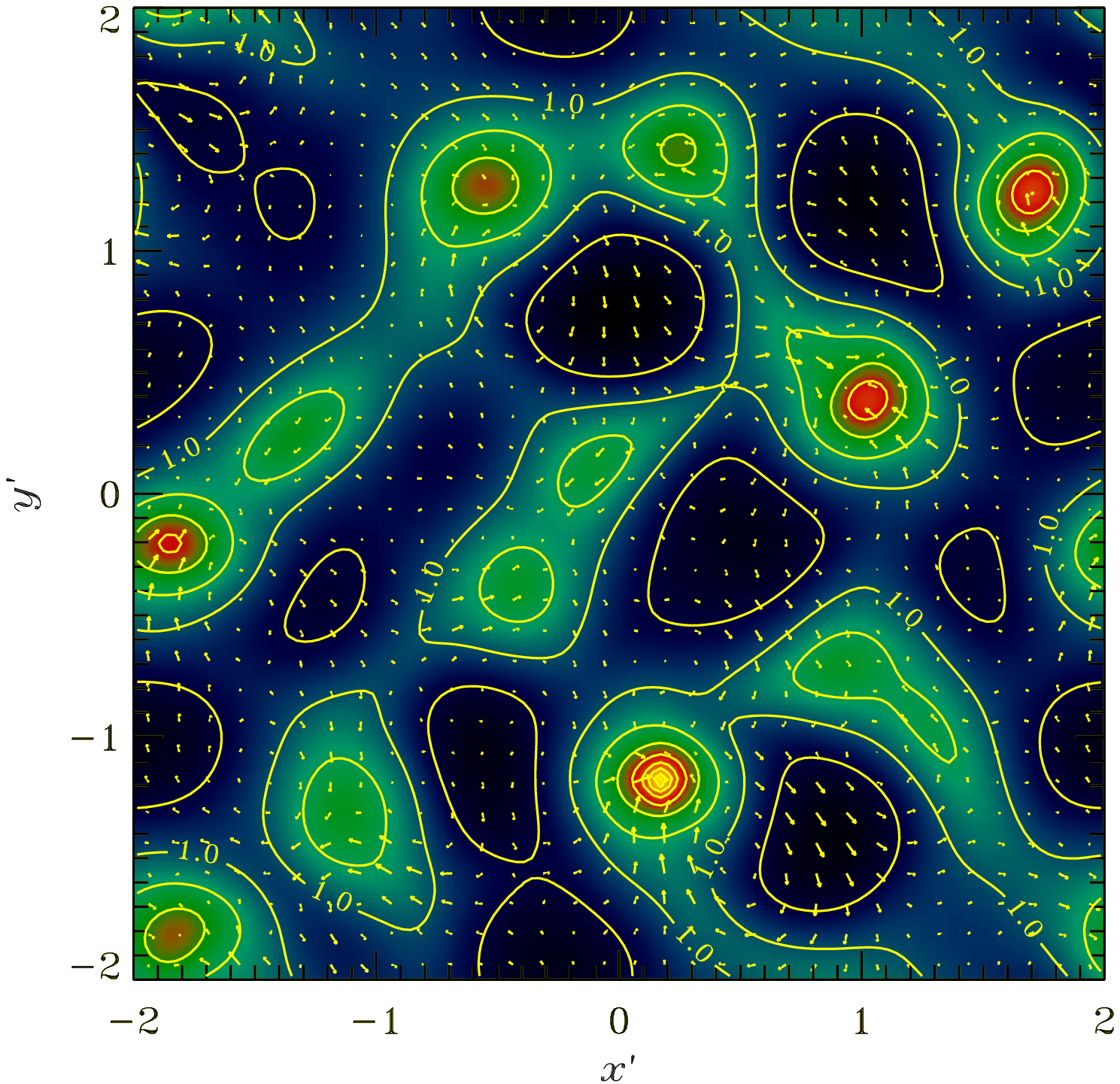,width=0.49\linewidth,clip=} &
\epsfig{file=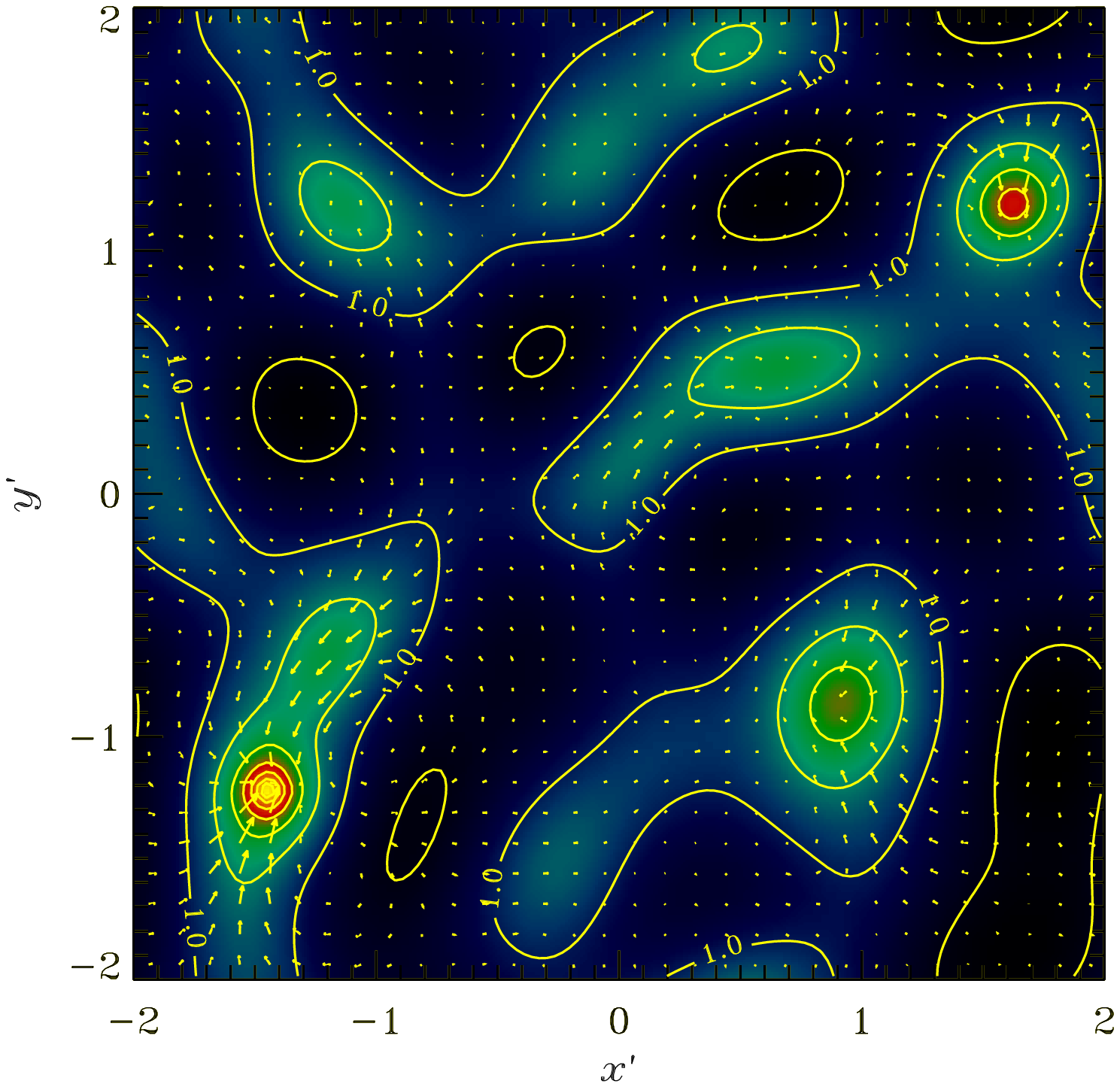,width=0.49\linewidth,clip=} \\
\epsfig{file=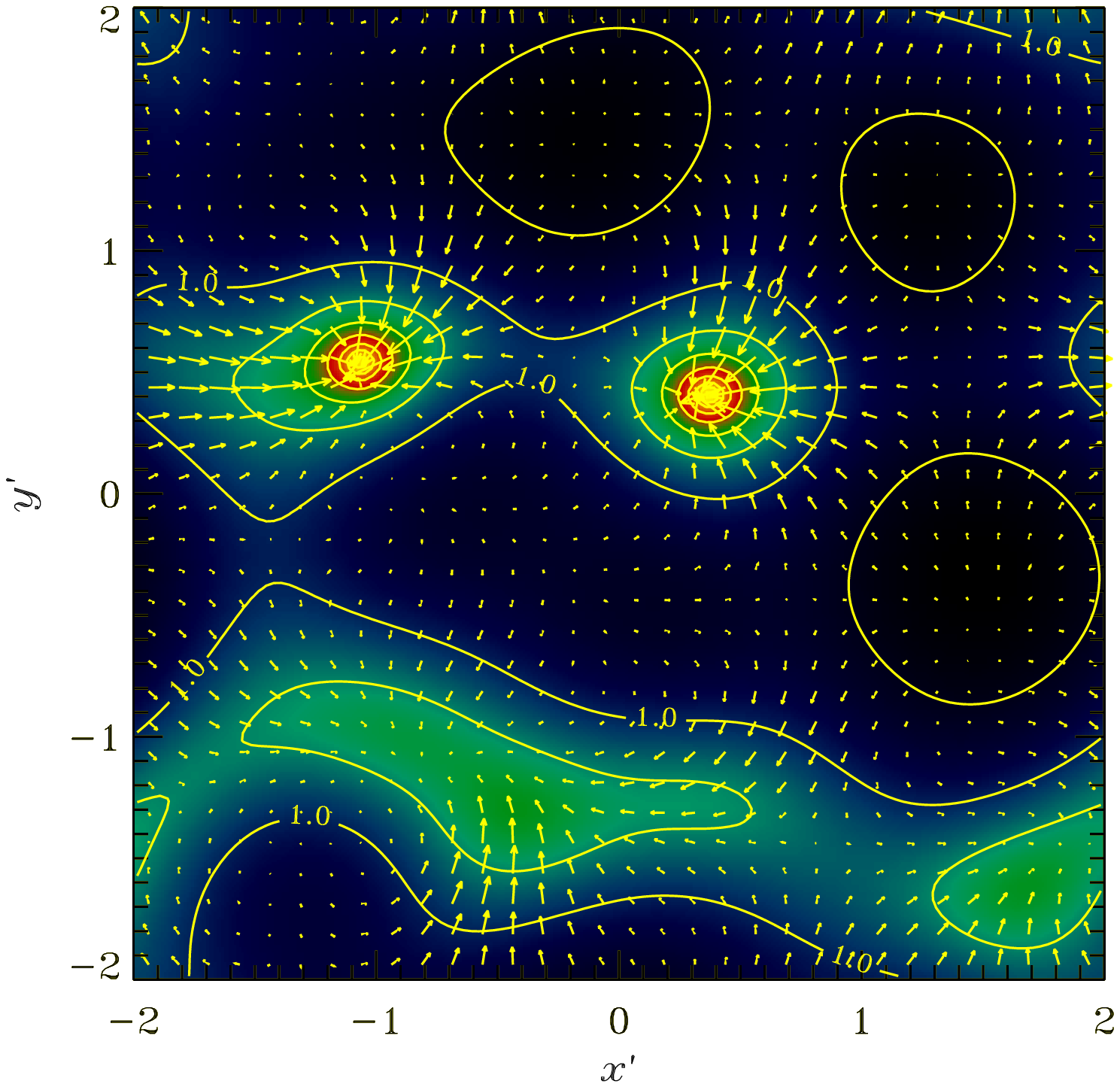,width=0.49\linewidth,clip=} &
\epsfig{file=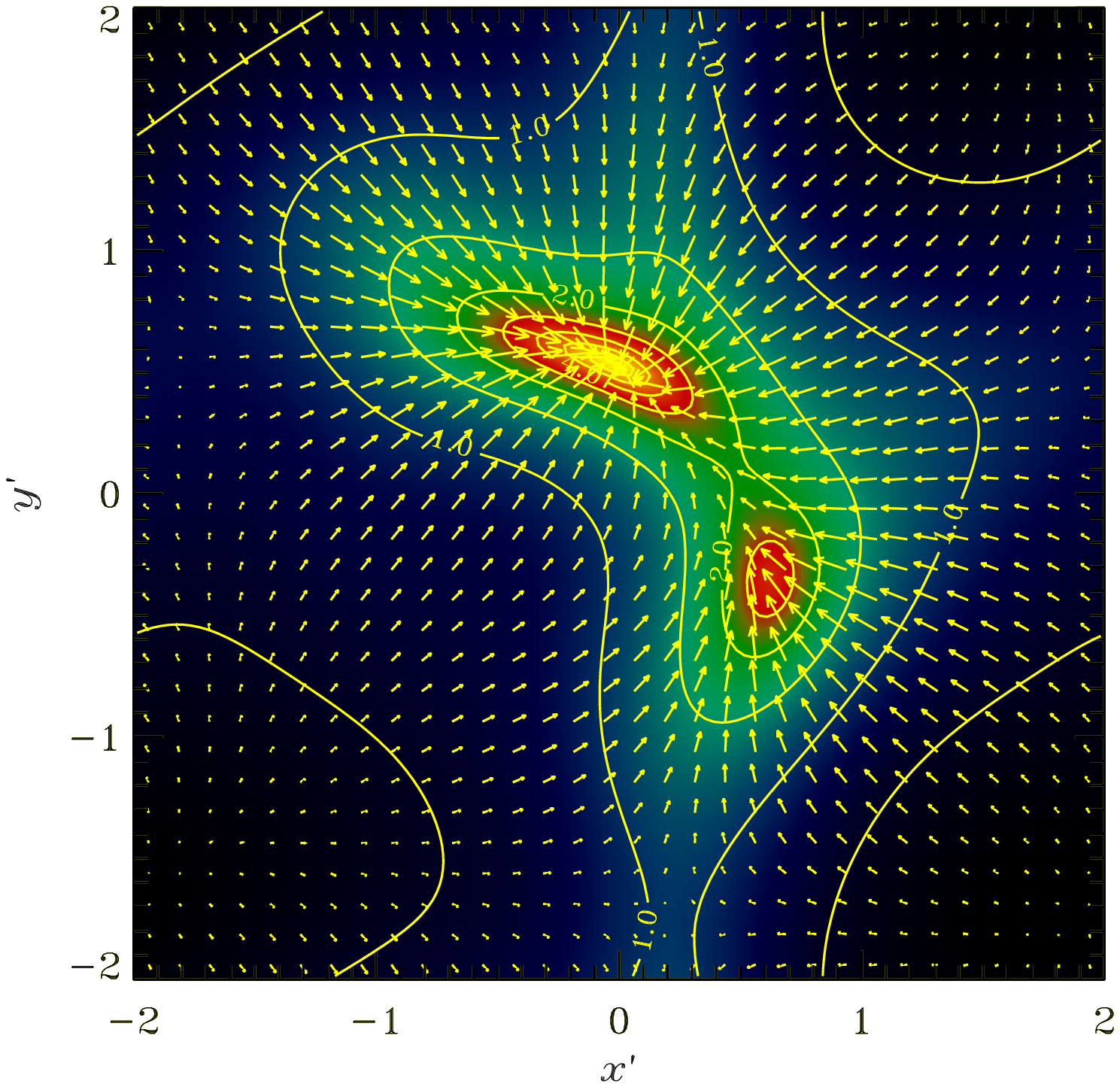,width=0.49\linewidth,clip=} \\
\epsfig{file=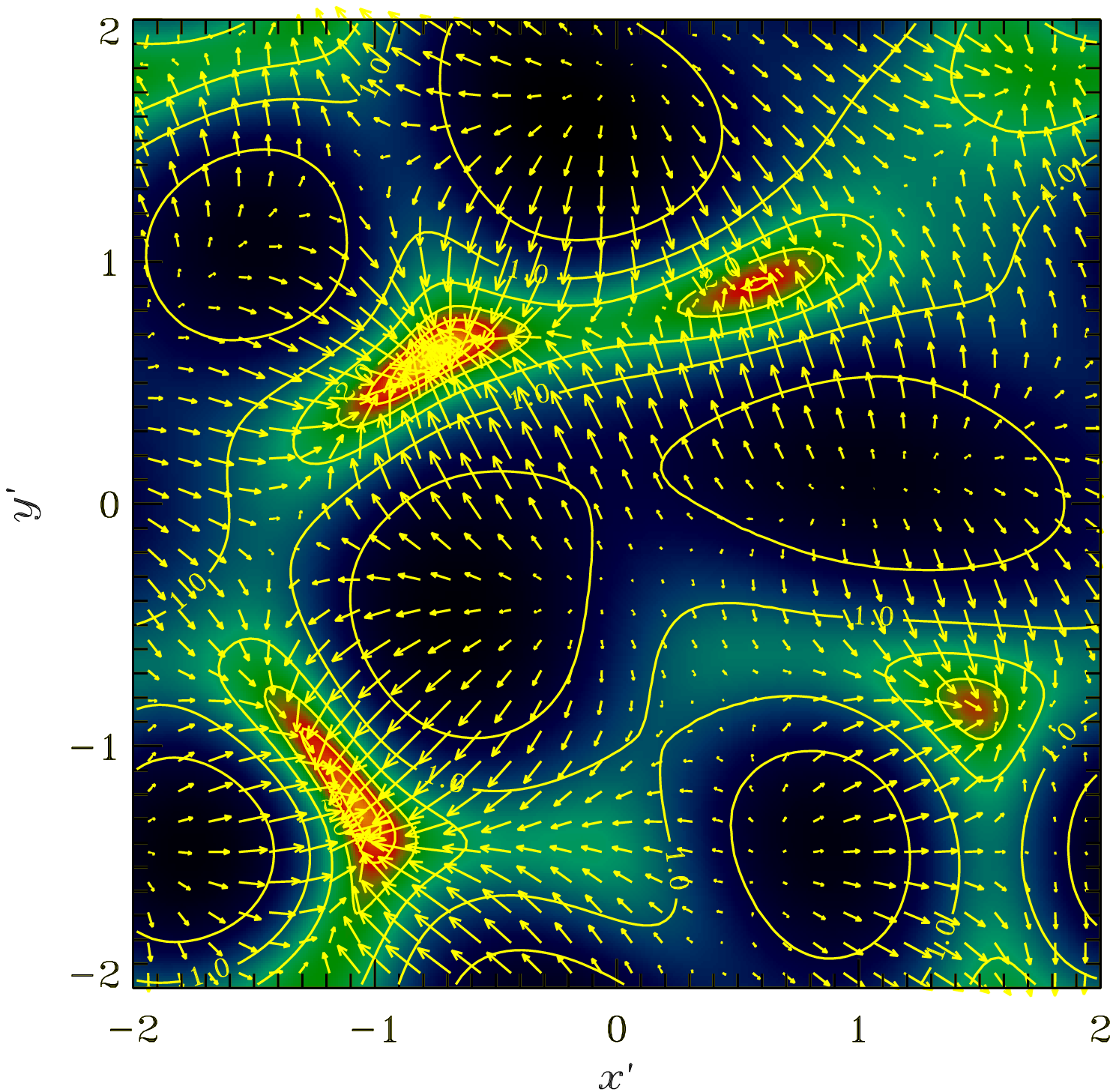,width=0.49\linewidth,clip=} &
\epsfig{file=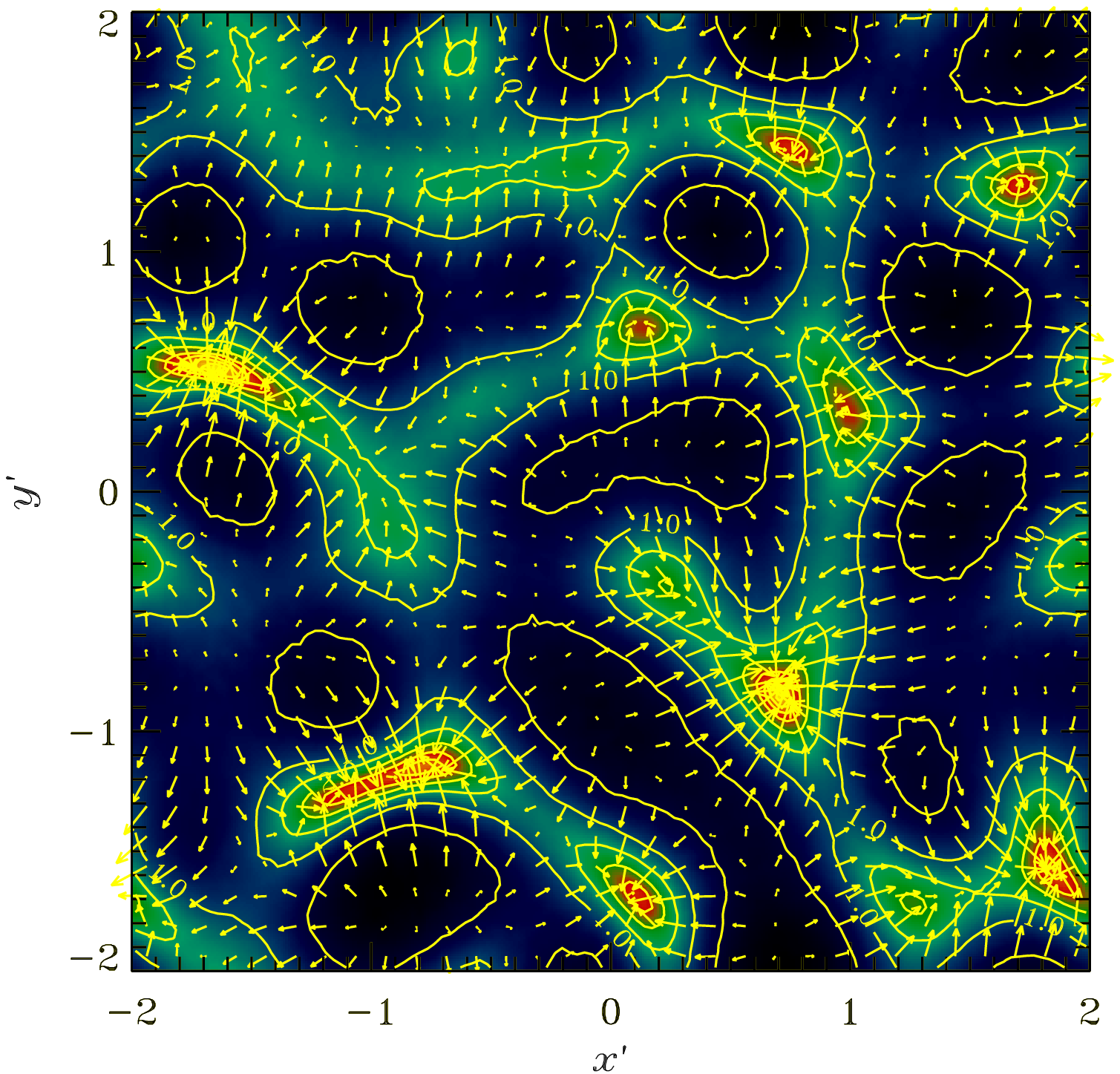,width=0.49\linewidth,clip=}
\end{tabular}
\caption{Image and contours of column density $\sign(x,y)/\signi$, 
and velocity vectors of neutrals, for six different models at the time 
that $\signmax/\signi = 10$. All models have $\tniitil=0.2$ and
$\Pexttil=0.1$. Top left: $\mui=0.5$. Top right: $\mui=0.8$. Middle 
left: $\mui=1.0$. Middle right: $\mui=1.1$. Bottom left: $\mui=2.0$. 
Bottom right: $\mui=10.0$. The color table 
is applied to the logarithm of column density and the contour lines 
represent values of $\sign/\signi$ spaced in 
multiplicative increments of $2^{1/2}$, having the values 
[0.7,1.0,1.4,2,2.8,4.0,...].
The horizontal or vertical distance between footpoints of velocity 
vectors corresponds to a speed $0.5 \, \cs$.
We use the normalized spatial coordinates 
$x' = x/\lammax$ and $y' = y/\lammax$, where $\lammax$ is the wavelength of
maximum growth rate in the nonmagnetic limit with $\Pext=0$.
}
\label{densimgs}
\efig

\bfig
\epsfig{file=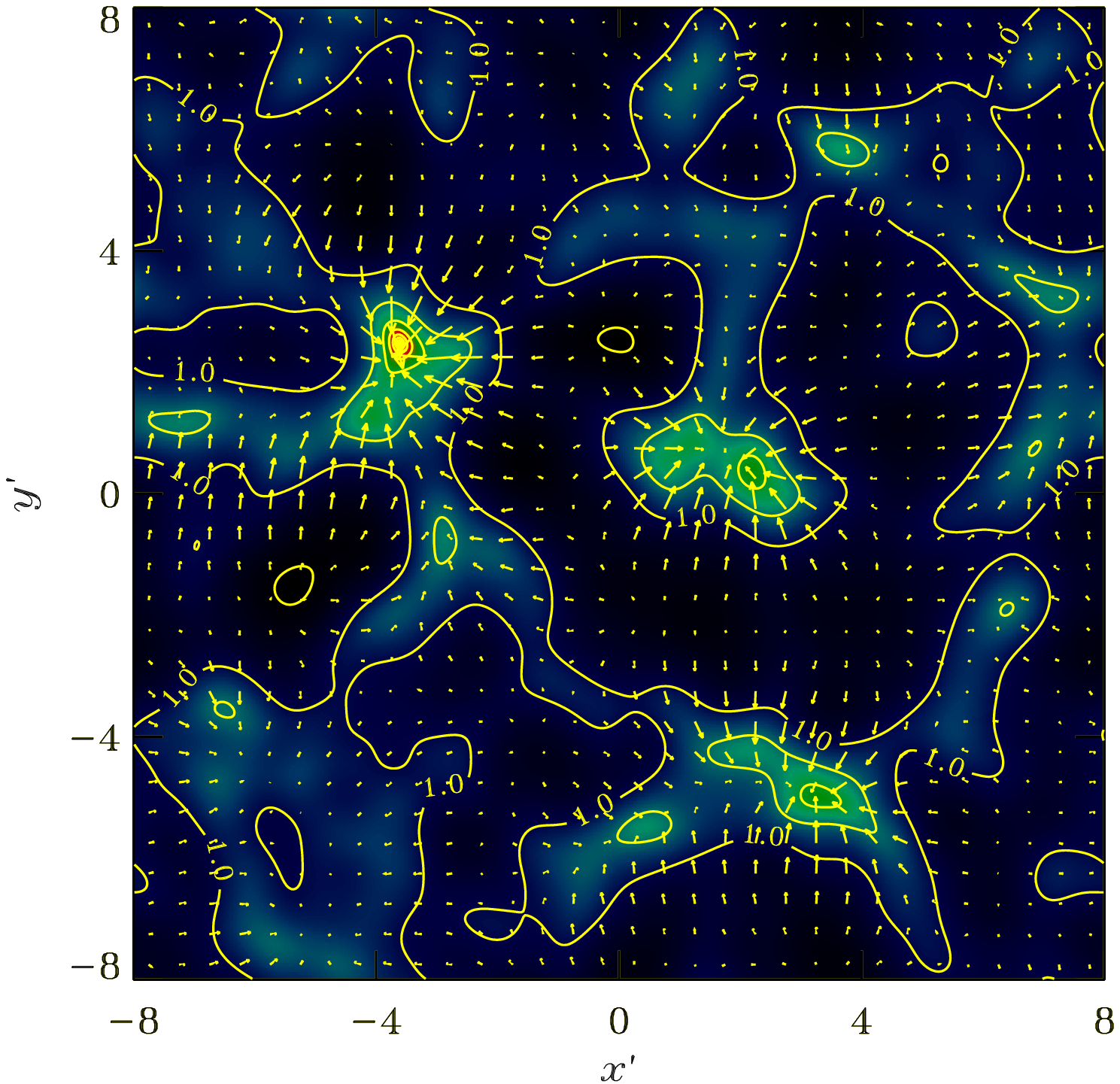,width=\linewidth}
\caption{Column density and velocity vectors as in Fig. \ref{densimgs} but
with model 4 parameters ($\mui=1.1$) that is run with four times larger 
computational region in each direction. The simulation
region consists of $512 \times 512$ zones.
The horizontal or vertical distance between footpoints of velocity 
vectors corresponds to a speed $0.5 \, \cs$.
}
\label{bigimage}
\efig

\bfig
\epsfig{file=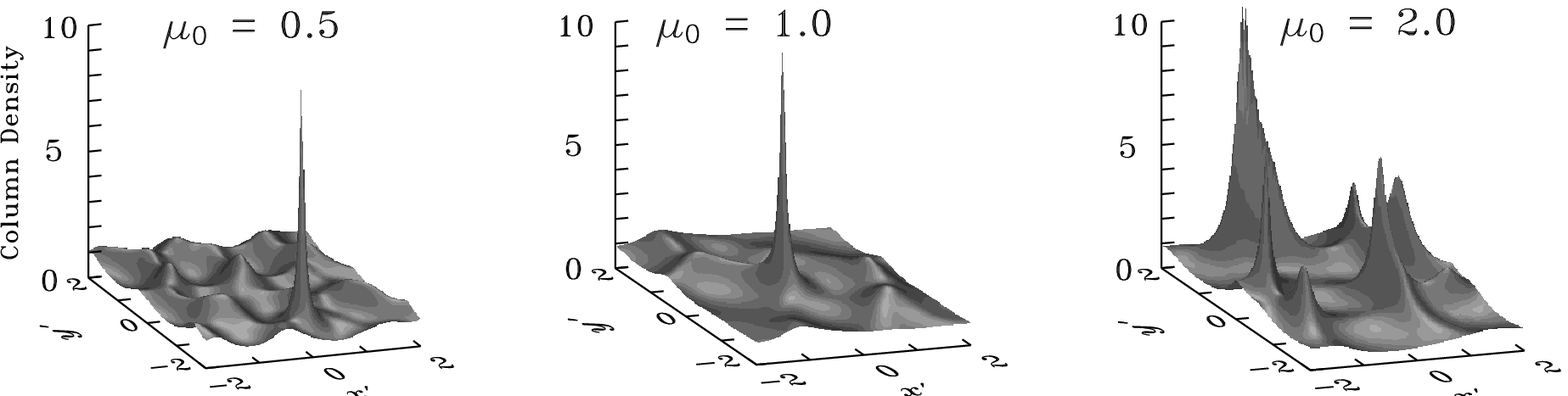,width=\linewidth}
\caption{Surface plot of column density $\sign(x,y)/\signi$ at the end of 
the simulation for models 1, 3, and 5 with (left to right) $\mui = 0.5,1.0,2.0$. 
Animations of the time evolution for each model are available online.}
\label{surfaceplots}
\efig

\subsubsection{Magnetic Field Lines}
\label{s:mfl}

For the force-free and current-free region above our model sheet, 
the three-dimensional structure of the magnetic field 
is obtained
by solving Laplace's equation for the scalar magnetic potential
$\Psim(x,y,z)$. The two-dimensional Fourier Transform of the
magnetic potential at any height $z$ above the sheet is related to its
known planar value $\Psi(x,y)=\Psim(x,y,z=0)$ via
\beq
\Psim(x,y,z) = \FTinv \left\{ \FT[\Psi(x,y)] \exp(-k_z\,|z|)\right\}.
\eeq
In the above expression, $\FT$ ($\FTinv$) represents the forward (inverse)
Fourier Transform in a 
two dimensional $(x,y)$ plane for a fixed $z$, $\Psi$ is obtained from
Eq. (\ref{magpot}), and $k_z = (k_x^2 + k_y^2)^{1/2}$.
We use the FFT technique to efficiently calculate $\Psim$ at any
height $z$, which then leads to the various components of the magnetic
field above the sheet via ${\bl B}-\Bref\zhat = -\nabla \Psim$
(see CB06 for a justification of this expression).

Fig.~\ref{fls} shows images of the final state column density for 
two models with 
$\mui=1.0$ and $2.0$, respectively. These are also independent realizations 
and differ in specific details from models presented earlier. 
Overlaid on the column density images are the magnetic field lines extending 
above the sheet. These are three-dimensional images of a sheet plus field
lines above, viewed from an angle of approximately $10^{\circ}$ from 
the direction of the background magnetic field. Clearly, the supercritical
model has more curvature in the field lines, as the contraction proceeds
primarily with field-line dragging. The critical model produces its
cores via a hybrid mode including both neutral-ion slip (ambipolar diffusion)
and field-line dragging, hence the the lesser amount of field line curvature.
The relative amounts of field line curvature in the cloud and within dense
cores are quantified by calculating the quantity $\theta = \tan^{-1} (|B_p|/\Beq)$, 
where $|B_p| = (B_x^2+B_y^2)^{1/2}$ is the magnitude of the planar magnetic
field at any location on the sheet-like cloud. Hence, $\theta$ is the angle that
a field line makes with the vertical direction at any location at the top
or bottom surface of the sheet. To quantify the differences in field line
bending from subcritical to transcritical to supercritical clouds,
we note that models 1, 3, and 5, with $\mui=(0.5,1.0,2.0)$, have average
values $\theta_{\rm av} = (1.7^{\circ}, 8.3^{\circ}, 18^{\circ})$, and maximum
values (probing the most evolved core in each simulation) 
$\theta_{\rm max} = (20^{\circ}, 30^{\circ}, 46^{\circ})$. 
For the model with $\mui=0.5$, $\theta_{\rm max}$ is comparable to 
that presented in Fig. 2 of \citet{Ciol}.

Our ability to model fragmentation with varying levels of magnetic
support and neutral-ion slip opens up the possibility of making a 
detailed comparison of observed hourglass morphologies of magnetic field
lines, where measured \citep[e.g.][]{Schl}, with theoretical models so that 
the curvature of field lines may be used as a proxy to measure the ambient magnetic 
field strength. 

\bfig
\centering
\begin{tabular}{cc}
\epsfig{file=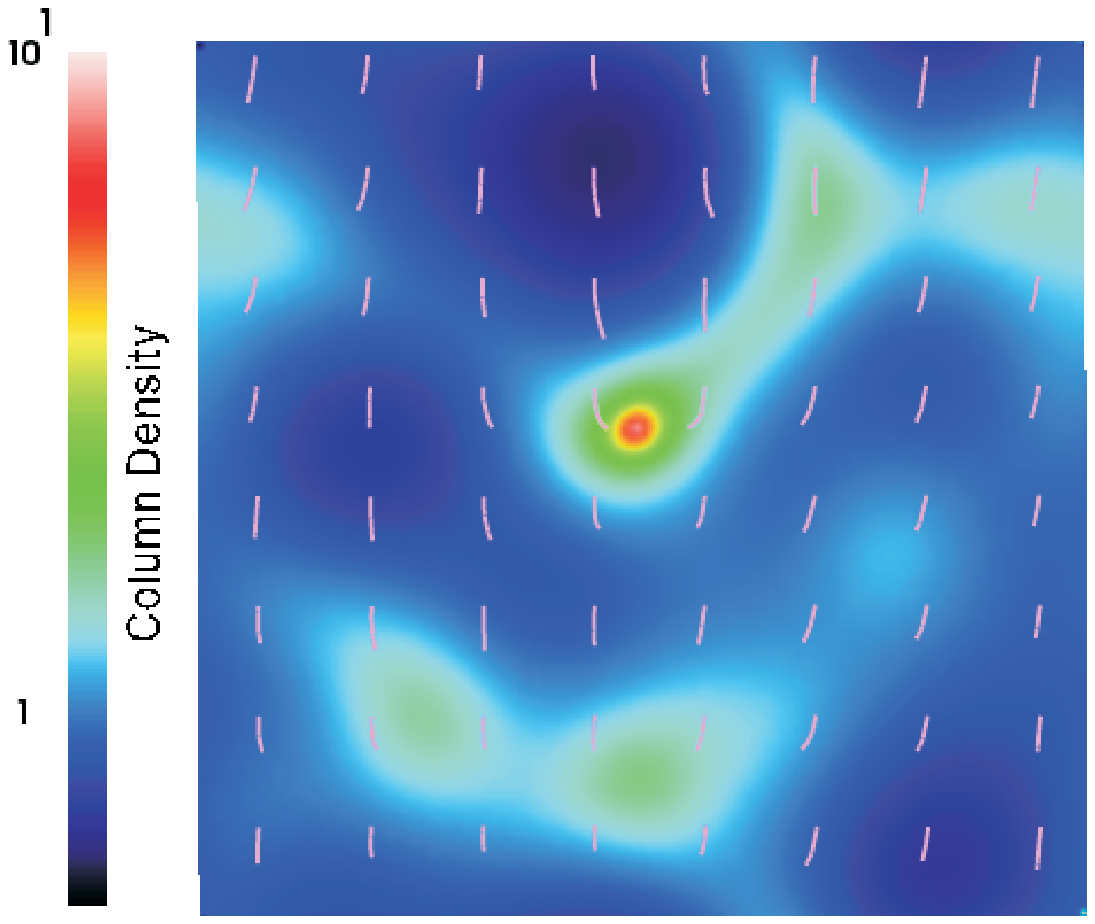,width=0.5\linewidth,clip=} &
\epsfig{file=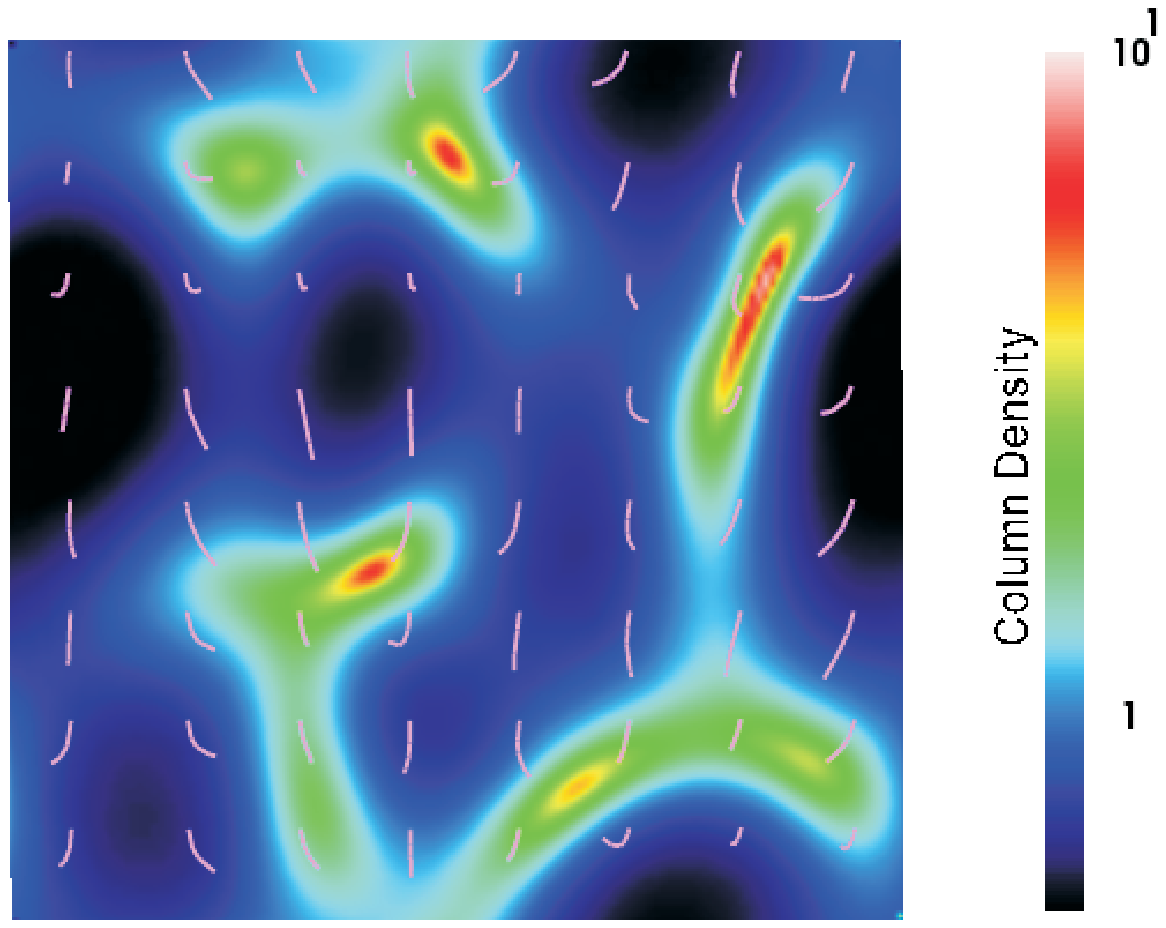,width=0.5\linewidth,clip=}
\end{tabular}
\caption{Image of gas column density $\sign(x,y)/\signi$ 
and superposed magnetic field lines for
models 3 and 5, with $\mui=1.0$ (left) and $\mui=2.0$ (right). The magnetic field
lines extend above the sheet, and the image and field lines are seen
from a viewing angle of about 10$^{\circ}$. Animations of the 
evolution of the column density are available online. The field lines
appear in the last frame of the animation.}
\label{fls}
\efig

\subsubsection{Shapes}
\label{s:shapes}

The core identification techniques
described in Section~\ref{s:techn} yield core sizes, shapes, and masses, whose average
values are tabulated in Table 2. Here in Fig.~\ref{shapehist} we present a more detailed 
histogram of core shape distributions for each of models 1 through 6. 
This isolates the effect of mass-to-flux ratio on core shapes. 
The statistics are generated by running each of model 1-3 and 5-6 for 
100 distinct realizations with the usual box size ($L=16\,\pi L_0,N=128$).
Model 4 is run 22 times due to the larger box size ($L=64\,\pi L_0,N=512$)
necessitated by the large fragmentation scale.
The number of cores generated are 
547, 367, 187, 126, 272, and 659, for models 1 through 6, respectively. The different 
numbers reflect the varying numbers of cores arising for each set of parameters
(see Fig. \ref{densimgs}) as well as the smaller number of runs for model 4. However, in each
case we have sufficient numbers to make inferences about the differences between
models.

Fig. \ref{shapehist} reveals in detail what is apparent by a visual inspection of 
Fig.~\ref{densimgs}. The core shape distribution contains many circular objects
(axis ratio $b/a \approx 1$) for subcritical clouds, but contains
progressively more elongated cores as $\mui$ increases.
All initially supercritical models have a peak axis ratio that is distinctly
non-circular. All objects are also flattened in the vertical direction and usually
have the shortest dimension along the magnetic field (see values of $\Zavg$ in
Table 2). 
The underlying physical explanation is that quasistatic formation
of cores (for $\mui \leq 1$) allows for growth in all directions more
equally, whereas dynamical gravity-dominated formation will strongly 
accentuate the anisotropies ($k_x \neq k_y$) that are present at the outset
\citep[see][]{Miya1,Miya2}.

The supercritical models have a mean axis ratio in the sheet
plane $\approx 0.5$, which is in rough agreement with the
observed mean {\it projected} axis ratio of dense cores \citep{Myer91}.
However, a deprojection of the observed axis ratios yields
intrinsic three-dimensional shapes that are inherently triaxial 
\citep{JBD01,JB02,Goodw,Tass},
with mean axis ratios $b/a \approx 0.9$ and $c/a \approx 0.4-0.5$.
Since the direction of the smallest axis ($c$) corresponds to our
preferred direction of flattening ($z$), the deprojected $b/a$ values 
can be compared directly with our models. We find reasonable agreement for
the subcritical models 1 and 2. The deprojected values of $c/a$ can also be
compared with our thin-sheet models, in which effectively $c/a = \Zavg/a = 
\Zavg \sqrt{\axisravg}/\savg$. There is reasonable agreement here again
for the subcritical models 1 and 2, as well as for the highly supercritical model 6,
which have, respectively, $c/a = 0.49, 0.43, ~{\rm and}~ 0.49$.
  
\bfig
\epsfig{file=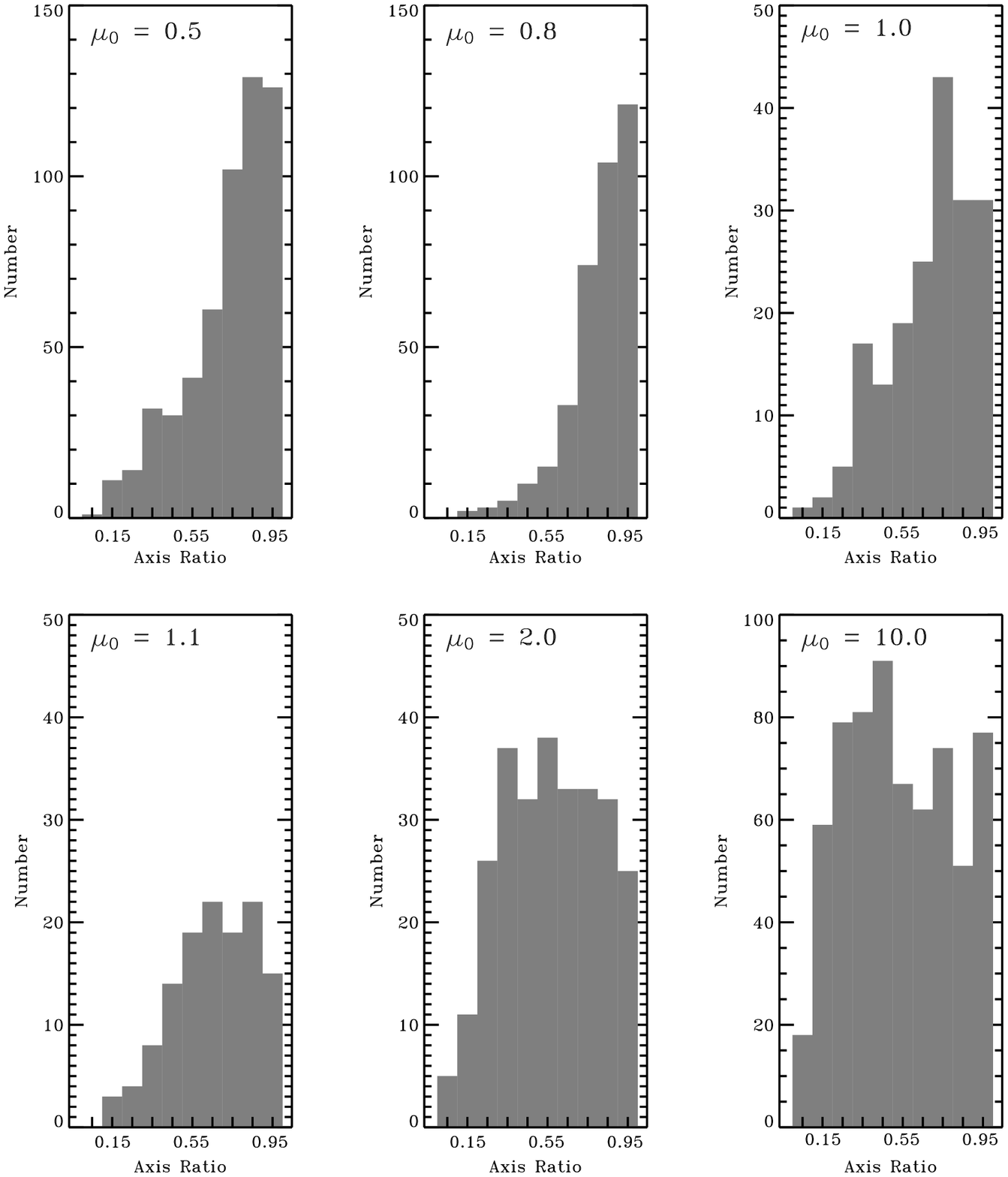,width=\linewidth,clip=}
\caption{Histograms of axis ratios $b/a$ of best fit ellipses to dense regions 
with $\sign/\signi \geq 2$, measured at the end of simulations with
$\mui =  0.5,0.8,1.0.1.1,2.0,10.0$ as labeled, corresponding to models 
1 through 6 in Table 2. Each figure is the result of
a compilation of results of a large number of simulations. The bin width is 0.1. }
\label{shapehist}
\efig

\subsubsection{Core Mass Distributions}
\label{s:cmd}

In Fig. \ref{masshist} we present histograms of core mass distributions generated
from the multiple runs of models 1 through 6, as 
described in Section~\ref{s:shapes}. Each core is defined as an enclosed region
with $\sign/\signi \geq 2$ that is present at the end of the simulation, when
$\signmax/\signi = 10$. For comparison with observations,
we have converted our calculated masses to
$\Msun$ using an assumed background number column density $\Nni =
10^{22} \cms$ and temperature $T=10$ K (see Eq. [\ref{M0}]).

We note that the variation of the peak masses (and the average masses tabulated
in Table 2) from one model to another are in qualitative agreement with 
the predictions of linear theory (Table 1). Furthermore, the $\mui=0.5$ and 
$\mui=10$ models generate very similar core mass distributions that are 
difficult to distinguish. This is not surprising since they have such similar
preferred fragmentation scales. 

The striking feature of each of the histograms is the very sharp descent
at masses greater than the peak of the distribution. Gravitational fragmentation
yields a very strong preferred mass scale.
The peak value itself is more ambiguous and can vary according to
the magnetic field strength, the background column density, the cloud temperature,
and the contour level we use to define the core.
In contrast, the slope on the low-mass side is much shallower. This is due to 
the capture of emerging cores at the end of any simulation. Many of those cores
are expected to grow in time and move over to the right by the time their 
peaks undergo runaway collapse and form a star. The distribution may be described as
relatively narrow and lognormal-like, but with a broader tail at the low-mass side
due to the temporal spread of core ages.

The steep decline of the of the mass distributions beyond the peak 
($d\log N/d \log M \approx -5$ is typical) 
in our study is in contrast to that observed for condensations in
cluster forming regions \citep[e.g.][]{Mott}, where $d\log N/d \log M \approx -1.5$ 
at high masses. Gravitational fragmentation under the conditions studied in
this paper and at the time snapshot chosen here yields a very strong preference
for a characteristic mass. We discuss possible mechanisms 
of broadening the mass distribution in Section~\ref{s:disc}. 

\bfig
\epsfig{file=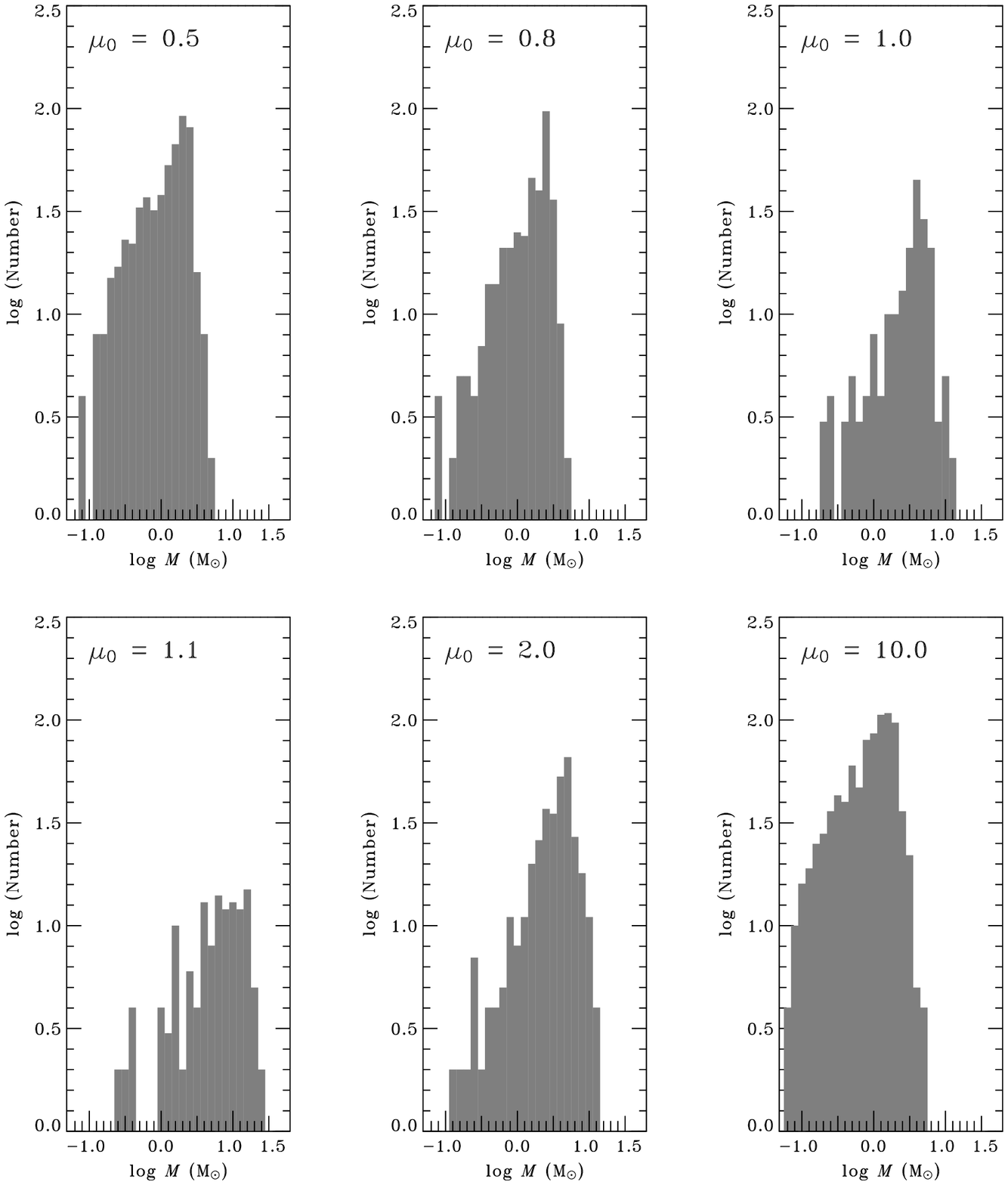,width=\linewidth,clip=}
\caption{Histograms of masses contained within regions with $\sign/\signi \geq 2$,
measured at the end of simulations with $\mui =  0.5,0.8,1.0.1.1,2.0,10.0$ as labeled. 
Each figure is the result of a compilation of results of a large number of 
simulations. The bin width is 0.1. }
\label{masshist}
\efig

\subsubsection{Supercritical Cores}

For clouds that start with subcritical or transcritical initial conditions,
there is available a more physical definition of a ``core'', i.e. a region that
is significantly supercritical and enclosed within a subcritical common cloud envelope.
Axisymmetric simulations of cores that evolve initially by ambipolar drift
have shown that the contraction becomes very rapid by the time that 
$\mu \approx 2$ in the central region, leaving behind a more slowly evolving
and essentially subcritical envelope \citep[see, e.g.][]{FM93,CM94,BM94}.

Fig. \ref{mtfimgs} shows images and contour maps of $\mu(x,y)$ at the 
end of the simulation
for models 1 and 3, respectively. The initially subcritical ($\mui=0.5$) model
1 has peaks in $\mu(x,y)$ coinciding with the major peaks in $\sigma(x,y)$
(see Fig. \ref{densimgs} upper left). However, note that the density condensations are 
either largely or even entirely subcritical ($\mu < 1$) at this stage.
This image shows that subcritical clouds can have observable density
enhancements which may still be partially or entirely subcritical, because they
are still in the process of ambipolar-drift-driven gravitational instability.
The image and contours for the initially critical ($\mui=1.0$) cloud shows
that the cloud naturally separates into supercritical and subcritical regions,
due to ambipolar diffusion and a fixed total magnetic flux threading the cloud.
The newly created supercritical regions are extended and typically contain more
than one density and mass-to-flux ratio peak within them. In this case, all
density peaks are associated with gas that has $\mu > 1$.

The presence of supercritical regions embedded within a common subcritical
envelope allows us to define cores in a more physical way than the previous
definition as regions with $\sign/\signi \geq 2$.
The latter is a somewhat arbitrary designation, as indeed are all observational
definitions of cores. However, the definition of supercritical cores has its own 
ambiguities, as a simple definition of regions with $\mu > 1$ yields extended
regions in model 3 with multiple density peaks. We find that a viable working 
definition is that a core is a region with $\mu > 1.3$. This isolates individual
density peaks in both models, and is consistent with the earlier axisymmetric
findings that a mass-to-flux ratio somewhat {\it above} the critical value
is necessary before rapid collapse and separation from the envelope becomes
apparent. For example, \citet{CM93} found that $\mu > 1.23$ was required for the 
absence of any available axisymmetric equilibrium state, using a similar
value of $\Pexttil$ as we do.
We compile data from 100 runs of each model with distinct random
realizations of the initial states and present the core mass distribution for
each of models 1 and 3 in Fig. \ref{scmasshist}. The conversion to dimensional 
masses is done
in the same manner as for Fig. \ref{masshist}. The resulting distributions 
have a peak mass
that is somewhat smaller than found using the different core definition used 
for Fig. \ref{masshist}. However, these distributions also have a very sharp decline 
at higher masses.

\bfig
\centering
\begin{tabular}{cc}
\epsfig{file=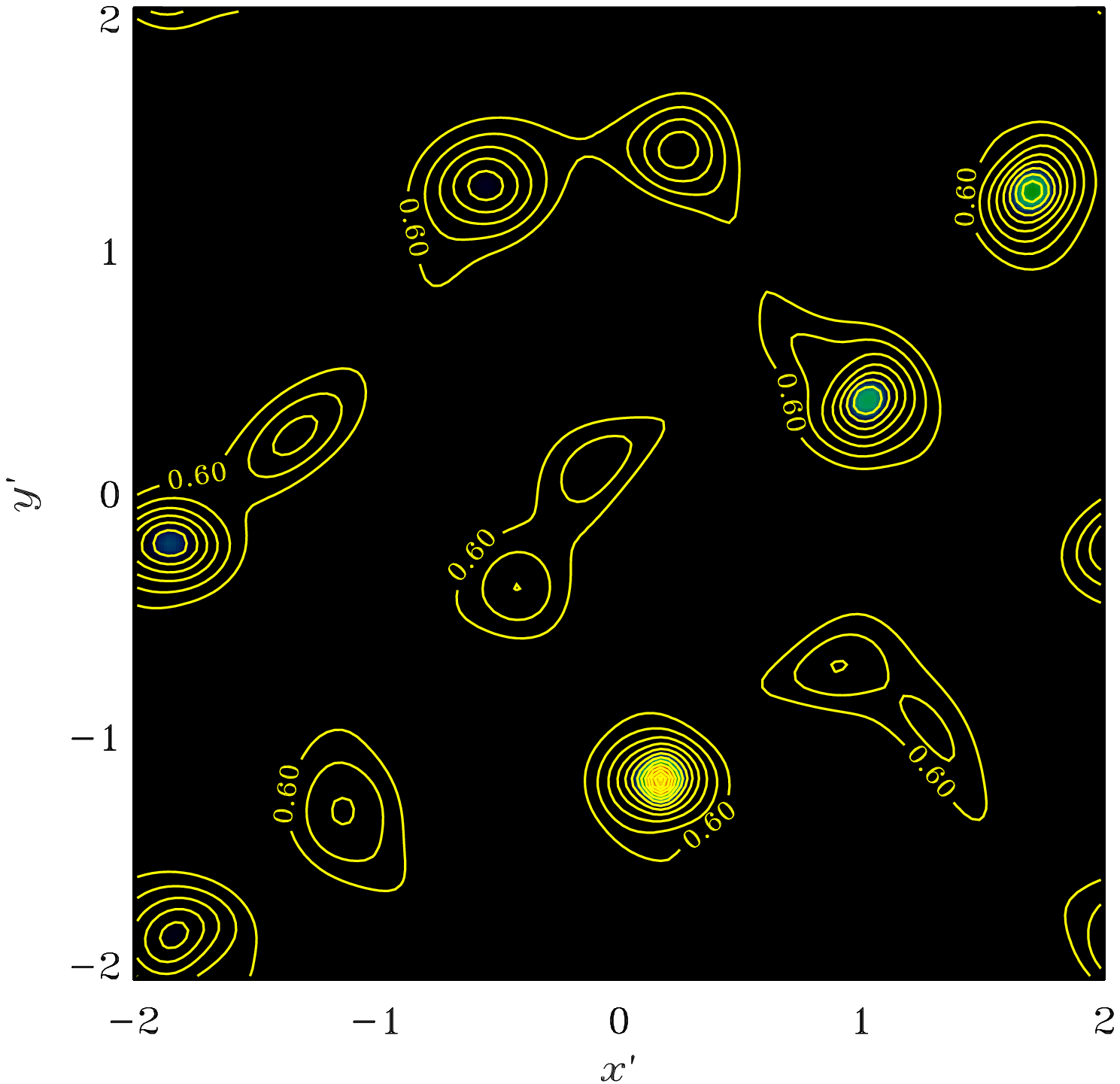,width=0.5\linewidth,clip=} &
\epsfig{file=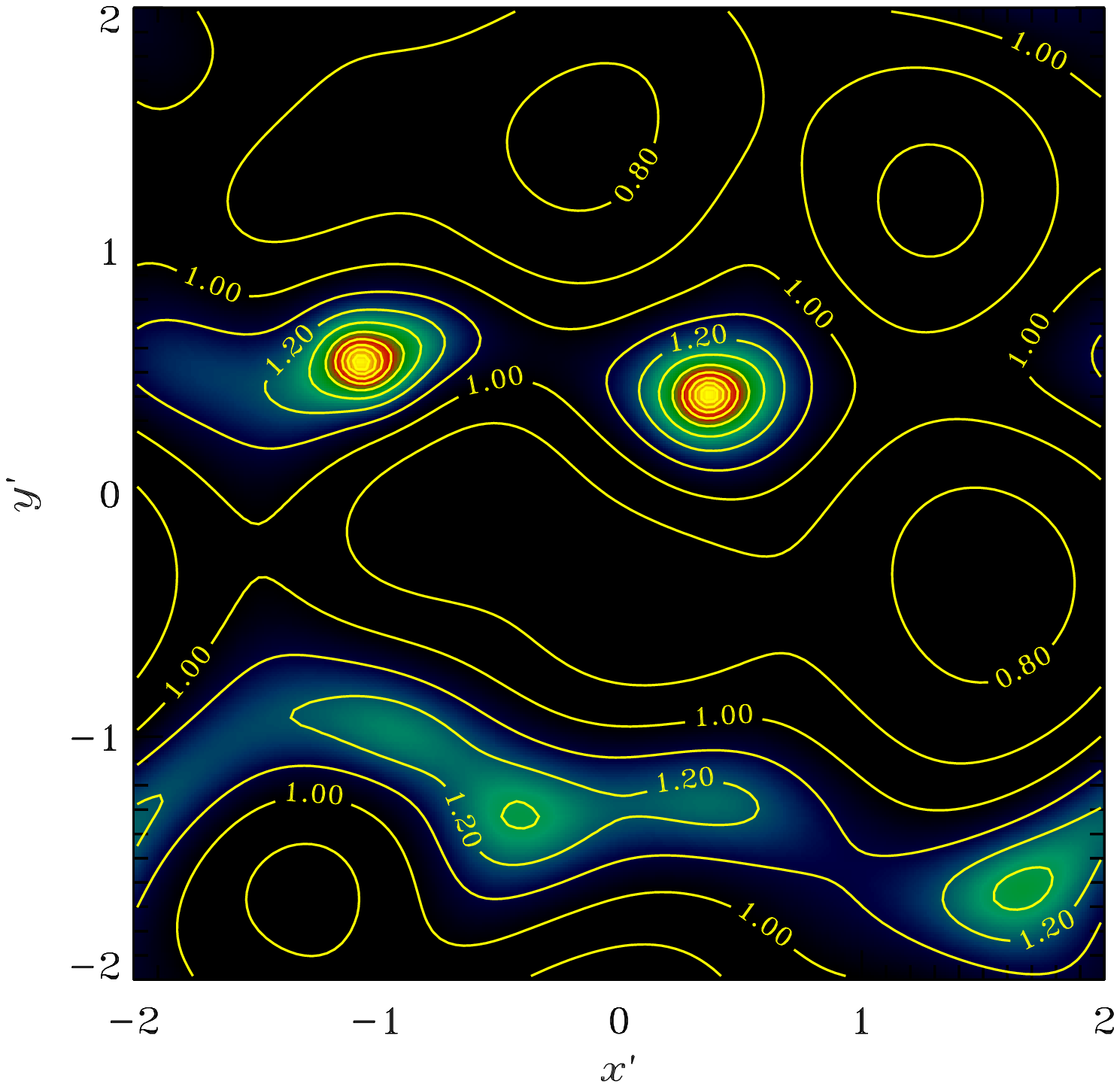,width=0.5\linewidth,clip=}
\end{tabular}
\caption{Image and contours of $\mu(x,y)$, the mass-to-flux ratio in units of the
critical value for collapse. Regions with $\mu >1$ are displayed with a color table,
while regions with $\mu <1$ are black. The contour lines are spaced in additive 
increments of 0.1. Left: Final snapshot of simulation with $\mui=0.5$. Right:
Final snapshot of simulation with $\mui=1.0$.}
\label{mtfimgs}
\efig

\bfig
\epsfig{file=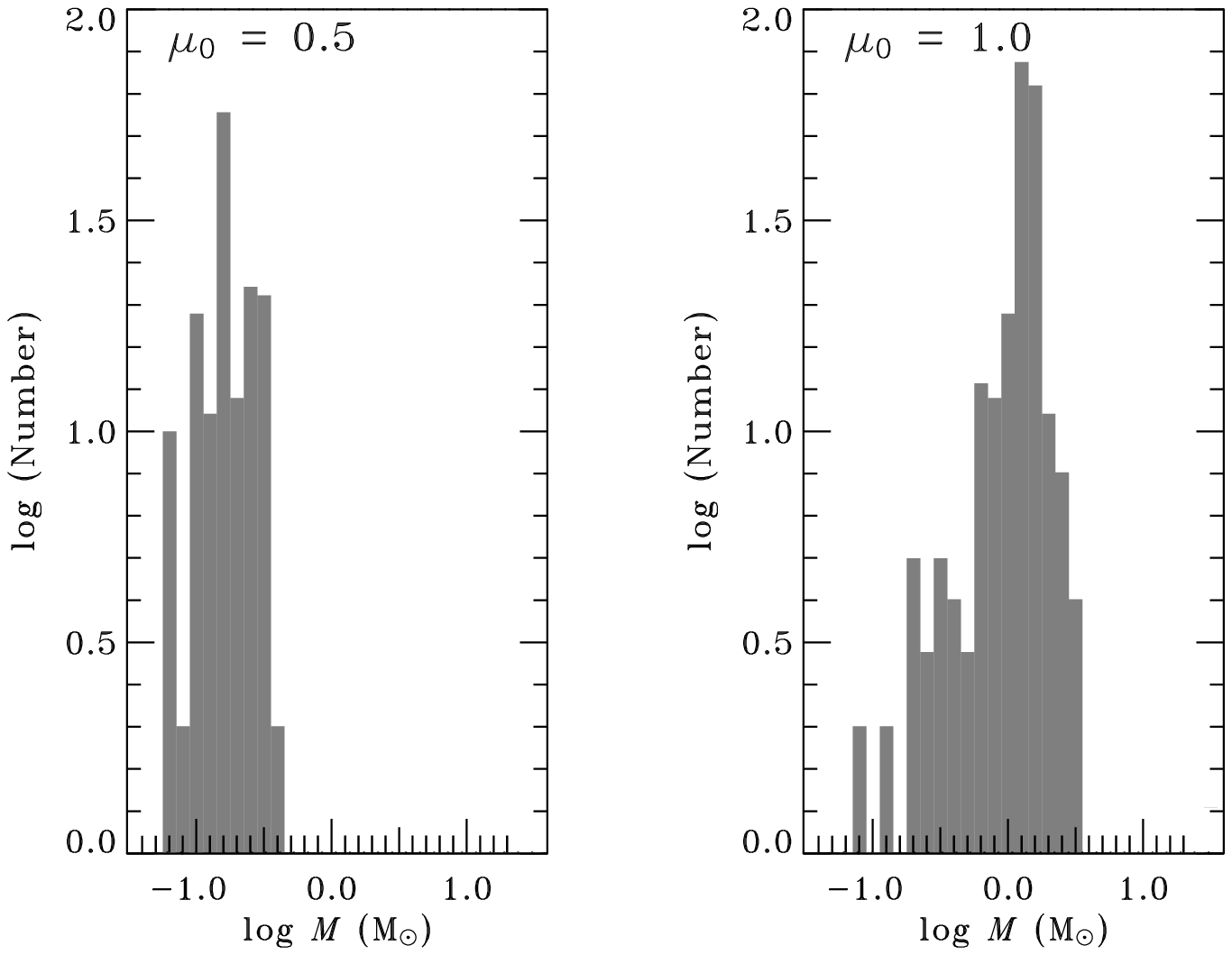,width=\linewidth,clip=}
\caption{Histogram of masses contained within regions that are significantly supercritical,
specifically $\mu > 1.3$, measured at the end of simulations with $\mui =  0.5$ and 
$\mui=1.0$. Each figure is the result of a compilation of results of a large number of
simulations. The bin width is 0.1.}
\label{scmasshist}
\efig

\subsection{The Effect of Varying $\tniitil$} 
\label{s:tni}

Eq. (\ref{ioniz}) shows that the ionization fraction $\xion$ at a given neutral density
$\nn$ increases (decreases) linearly as $\tniitil$ decreases (increases).
We investigate the effect of decreasing and increasing $\tniitil$
by a factor of two from its standard value 
(which can be accomplished by changing the factor ${\cal K}$ in Eq. [\ref{rhoieq}])
in models 7 and 8, respectively. The other two parameters
are kept fixed at $\mui=1.0$ and $\Pexttil=0.1$. The 
characteristic growth times of instability $\taugm$ scale approximately
$\propto \tniitil^{-1} \propto x_{\rm i,0}$ (see also Table 1), where $x_{\rm i,0}$ is the
ionization fraction at the background number density $\nni$.
Models 7, 3, and 8 have $\tniitil=(0.1,0.2,0.4)$, $\taugm = (27.5,14.3,7.9) \times t_0$, 
and $\lamgm = (28.5, 24.3, 20.0) \times L_0$, respectively.
Furthermore, Table 2 reveals that the time to runaway collapse
($\signmax/\signi \geq 10$) is $\approx 10\, \taugm$ when starting from small-amplitude
white noise perturbations, as generally found in our parameter study.
This means that our high ionization-fraction model 7 has the largest value of
$t_{\rm run}(=261\,t_0)$ in our parameter study. This also leads to the largest
age spread of cores in any of our simulations. This is measured by the fact that a typical
simulation, when run (for compiling statistics) in a large box with $L=64\,\pi,N=512$,
has many cores that are just beginning to emerge when the first core goes into a runaway 
collapse. Hence, our value for $\lamavg$ is calculated with a lower core threshold
$\sign/\signi \geq \sqrt{2}$ for this one model. Our analysis reveals that
the fragmentation 
scales in the nonlinear phase are indeed comparable amongst models 3, 7, and 8, and
in good agreement with the linear theory prediction.
We conclude that the effect of varying ionization (for a fixed $\mui$) within the range studied is 
primarily in the {\em rate} of evolution.

Fig.~\ref{tnidensimgs} shows images and contours of the density, as well as
velocity vectors, for models 7 and 8.   
The time to reach runaway collapse is about four times longer
for model 7 than for model 8, consistent with its value of $\tniitil$ 
being four times smaller. Model 8 has $\tniitil = 0.4$, hence poorer 
neutral-ion coupling and therefore reduced magnetic support. This results
in slightly
greater infall speeds, slightly larger number of fragments, and cores 
which are slightly more elongated. These are all consistent with its more
dynamical evolution.



\bfig
\centering
\begin{tabular}{cc}
\epsfig{file=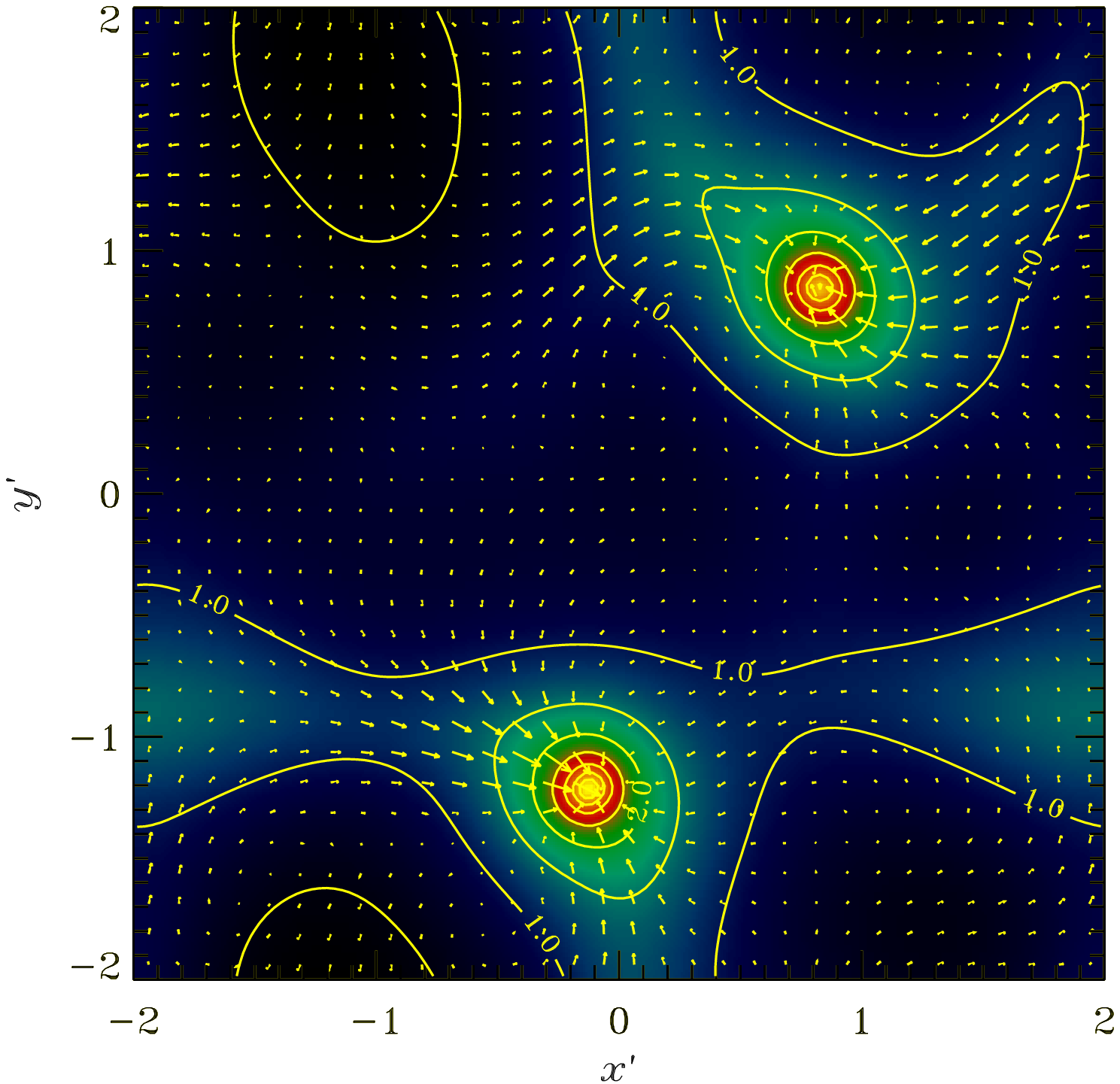,width=0.5\linewidth,clip=} &
\epsfig{file=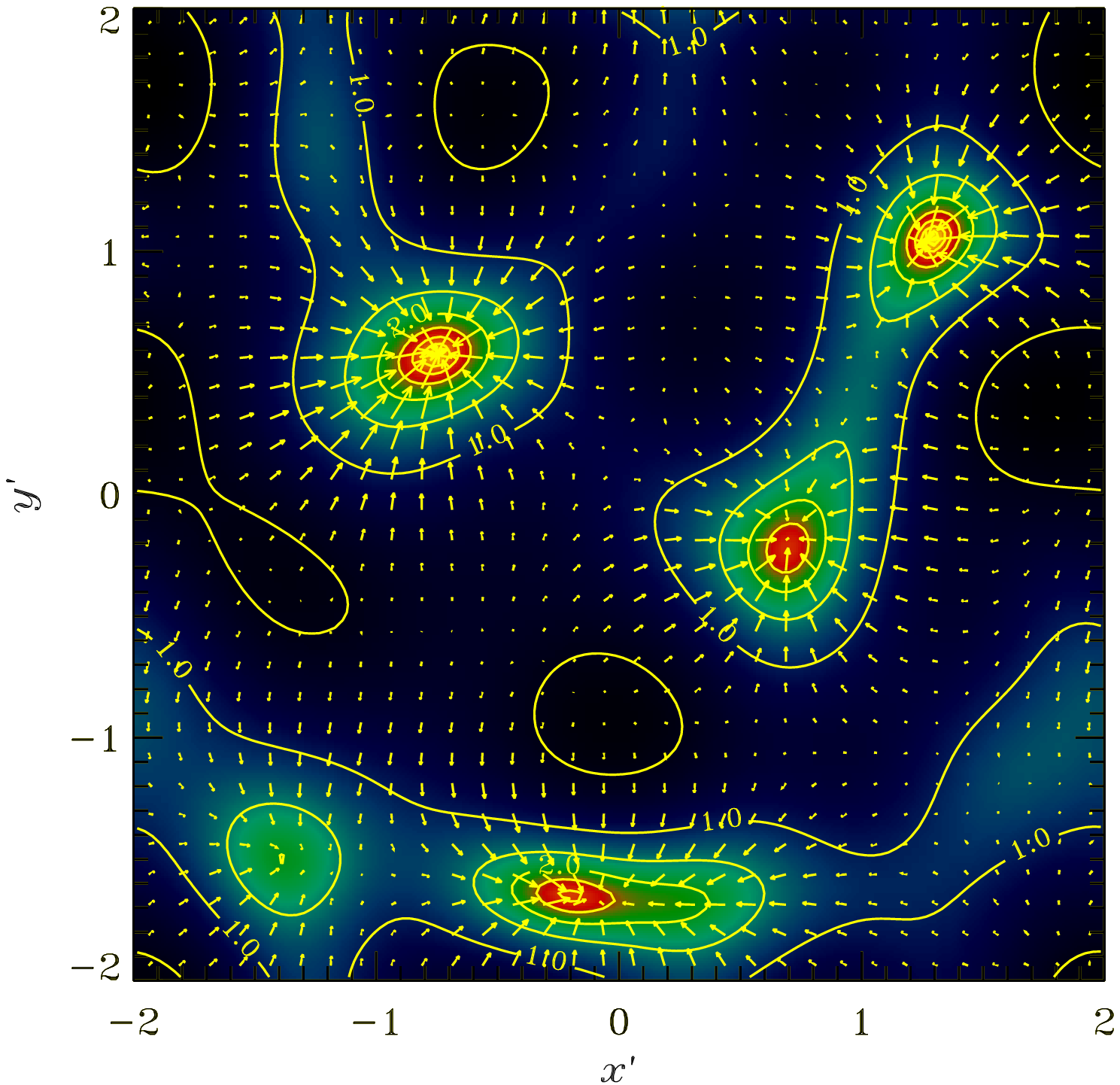,width=0.5\linewidth,clip=}
\end{tabular}
\caption{Column density and velocity vectors as in Fig. \ref{densimgs}, but for models
with $\tniitil = 0.1$ (left) and $\tniitil=0.4$ (right). Both models have
$\mui=1.0$ and $\Pexttil=0.1$.}
\label{tnidensimgs}
\efig

\subsection{The Effect of Varying $\Pexttil$} 
\label{s:pext}

Our models with $\Pexttil=10$ may represent the effect of pressured environments 
such as sheets brought together by the presence of shocked gas 
(e.g. stellar winds or supernovae) and being embedded in or adjoining 
an H {\sc II} region. These models could represent an example of 
``induced'' star formation in a 
manner related but not equivalent to that of an initial turbulent flow 
with high ram pressure. 

Models 9 and 10 both develop extreme clustering in comparison to the other models. 
Table 2 shows that the fragmentation scales are about 1/3 to 1/4 of that for the 
corresponding models with the same mass-to-flux ratio but small $\Pexttil$.
The fragments grow initially through a pressure-driven mode and the spacing
is in excellent agreement with the predictions of linear theory (Fig. 
\ref{scales} and Table 1).
However, our results show that the nonlinear instability does develop into a 
gravitationally-driven runaway collapse.
The maximum speeds are still subsonic at the end of our simulation, due to 
the relatively weak gravitational influence of each compact core. As well as
having the smallest fragmentation scales, these models also have the shortest
time scales to runaway collapse. For the fiducial $\Nni = 10^{22} \, \cms$,
models 9 ($\mui=0.5$) and 10 ($\mui=1.0$) have values of $t_{\rm run} =
(1.6 \Myr, 0.81 \Myr)$ and $\lamavg = (5800 \AU, 7700 \AU)$, respectively.
Both sets of numbers are considerably smaller than for the models 1 and 3,
which have corresponding values of $\mui$ but $\Pexttil=0.1$.
Fig.~\ref{pext10densimg} shows the clustering properties of model 10 at the end of the simulation, 
which is very similar to the corresponding image for model 9 (not shown).
This fragmentation model clearly produces a much richer cluster than in the
relatively unpressured environments presented earlier. Velocity vectors are not
shown in this image due to confusion arising from infall onto so many peaks.
A careful inspection of the image reveals a variety of core spacings and
sizes at this stage of evolution. Interestingly, the average core spacing
$\lamavg = 5.3 L_0$ is in very good agreement with the preferred 
wavelength in linear theory, 
$\lamgm = 5.0 L_0$. These length scales are well resolved in our 
simulations. However, the average core mass $\Mavg$ significantly exceeds the 
linear theory value $\mgm$. Only models 9 and 10 show such a large
discrepancy between these values. We attribute it 
to the very small sizes of the cores (see $\savg$ values in Table 2)
in these simulations. 
This means that the cores themselves are barely resolved and the mass
estimates should be taken as approximate values that likely represent
upper limits.

\bfig
\epsfig{file=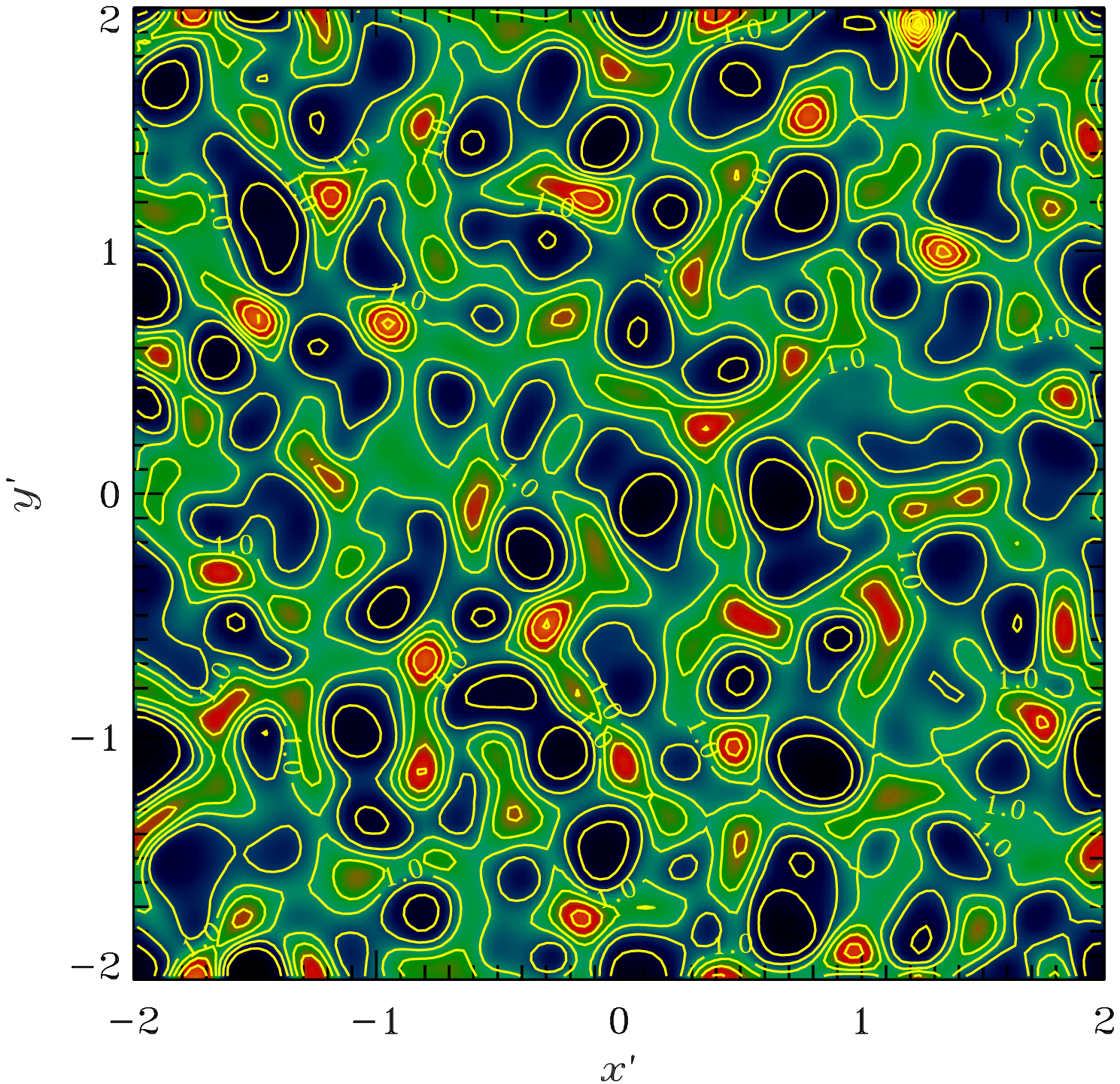,width=\linewidth,clip=}
\caption{Column density as in Fig. \ref{densimgs}, but for a model
with $\Pexttil = 10$. Other parameters of this model are 
$\mui=1.0$ and $\tniitil=0.2$.}
\label{pext10densimg}
\efig

\section{Discussion}
\label{s:disc}

Our simulations of the nonlinear development of gravitational instability
under the influence of magnetic fields and ambipolar diffusion start
from a background state of uniform column density and magnetic field
strength. Small-amplitude white-noise perturbations initiate the evolution and
eventually lead to the nonlinear growth of fragments. Averaging over a large number
of simulations reveals that the average spacing of nonlinearly developed 
cores is essentially that predicted from the preferred fragmentation
scales in linear perturbation theory (CB06). However, the time to
reach fully developed runaway collapse is up to ten times longer than
that of the eigenmode with minimum growth time $\taugm$.
The quantity $\taugm$ itself varies from $\approx Z_0/\cs$ 
(essentially the free-fall time $\approx 1/\sqrt{G\rhoni}$ for unpressured
sheets) for highly supercritical
models to $\approx 10 Z_0/\cs$ for highly subcritical models (for a typical neutral-ion
coupling level).
The times to reach runaway collapse vary widely amongst
models with different mass-to-flux ratios, ionization fractions, and
external pressures. For a cloud with $\Nni=10^{22}\cms$ and $T=10\K$,
the times to reach runaway growth of the
first core ranges from 0.45 Myr to 9.53 Myr (see Table 2).    
Since our simulations start from a flat density background, these times
represent {\it upper limits} to the time that fragmentation might take
for each set of parameters. However, an advantage
of the uniform background density is that it allows for a self-consistent
modeling of the entire core formation process, without questions
about the origin of initially peaked density distributions used in earlier
axisymmetric calculations \citep[e.g.][]{CM93,BM94}. 


In a medium with initial nonlinear perturbations,
the time scales for all sets of parameters are indeed likely to be shorter.
However, we believe that our calculated time scales are
relevant if the corresponding dimensional values are obtained from higher 
starting values of column density brought about in certain regions by 
pre-existing (including turbulent) flows.  
For example, the Taurus molecular cloud has an overall background number column 
density $N \approx (1-2) \times 10^{21} \cms$ but also contains embedded
dark clouds with $N \approx 5 \times 10^{21} \cms$, such as HCl 2 and L1495, 
within which there are small clusters of $\sim 10-20$ YSO's and also 
many dense cores \citep[see][]{Gome,Onis,Gold}. 
Our periodic model may be applied to such dark clouds that may themselves
have been brought together by nonlinear flows originating in the larger cloud,
external triggers, or an earlier phase of gravitational fragmentation. 
A second example application of our periodic model may be 
within the L1688 dark cloud in Ophiuchus, which has 
$N \approx 10^{22}\,\cms$ \citep{Mott}. 
The mean spacing of fragments in these
two regions varies \citep{Andr00}, with core edges measured to be at radii
$\la 5000\, \AU$ in L1688 but at $\la 20000 \, \AU$ in the Taurus dark clouds
\citep{Andr00}. While some of the difference in spacing may be due to the 
different background column densities, other important aspects of spatial and
kinematic structure may also arise due to 
different values of $\mui$ and $\Pexttil$ (and $\tniitil$ to a lesser extent), 
as demonstrated in this paper. 

Unlike a uniform density three-dimensional medium, our thin sheet actually
has a preferred scale for gravitational fragmentation with a unique value
for any given set of initial dimensionless parameters. Since our simulation
region is always safely larger than this fragmentation scale, it is
unlikely that
the size of our system (in the $x$- and $y$-directions) influences 
the final outcome, as measured by
fragment spacings, time scales to runaway collapse, and core mass distributions, for
example. This is supported by our tests with runs at quadruple the size of
the standard simulations. 
Incidentally, this is not the case for three-dimensional periodic box simulations,
in which the fastest growing mode of gravitational instability is always that
of the box size.
We believe that a stratified medium is also a more physical and realistic 
starting point
than a uniform three-dimensional medium. This is because the formation process 
of molecular 
clouds, or magnetic fields and nonlinear flows within it, will tend to set up compressed 
regions with 
a characteristic scale similar to the half-thickness $Z_0 \approx
\csq/(\pi G \signi)$ of our adopted background reference
sheet geometry. That scale is related to the Jeans scale and effectively
determines the preferred fragmentation scale, as modified by magnetic field
strength, ionization fraction, and external pressure.




\bfig
\epsfig{file=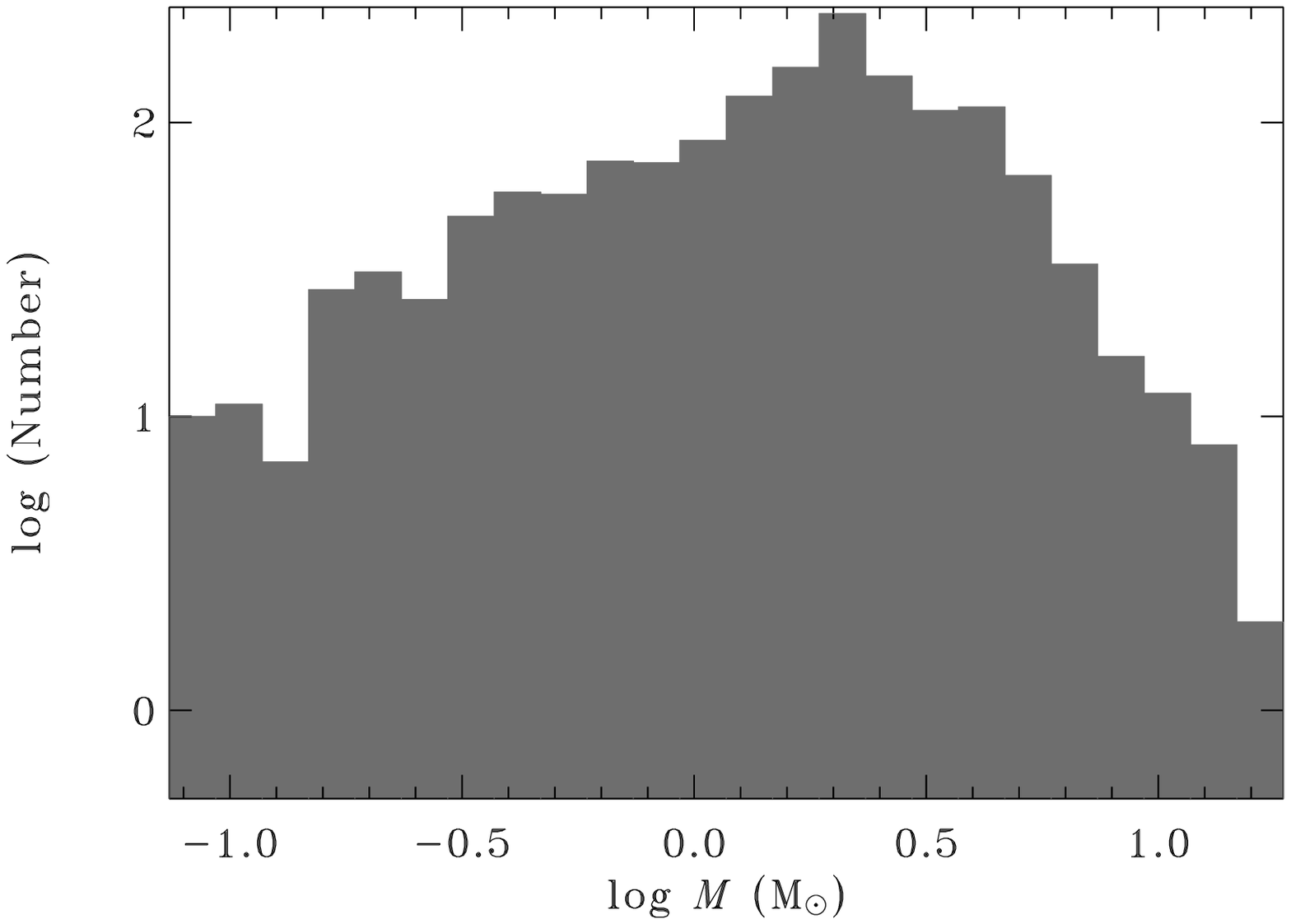,width=\linewidth,clip=}
\caption{Histogram of masses from models with $\mui=0.5,0.8.1.0,1.1,2.0$ and
$\tniitil=0.2,\Pexttil=0.1$. A weighted average of core masses is taken,
so that an equal simulated area is assigned to each of models $1-5$.
The cores are the same as the ones in the first five panels of
Fig.~\ref{masshist}. The bin width is 0.1.
}
\label{totalmasshist}
\efig

The core mass distributions that we have compiled from our simulations
are narrowly-peaked and do not match the broader distributions commonly
observed in cluster-forming regions \citep[e.g.][]{Mott}.
What is the missing physics that can explain the discrepancy?
The explanation that is closest to the spirit of our
models is that real molecular clouds start their lives with an
inhomogeneous distribution of physical quantities, including the mass-to-flux
ratio. We expect that the values of $\mu$ in a molecular cloud
may fall within the observationally-established range of $[0.5,2]$, i.e. within
a factor of two of the critical value in each direction. In that case, 
we can make a simple estimate of a core mass distribution by adding up the
histograms for the five models with $\mui=[0.5,0.8.1.0,1.1,2.0]$ shown
in Fig.~\ref{masshist}. Since the model with $\mui=1.1$ is run fewer times but with a
larger box size, we sample the number of cores necessary to give equal area
weighting with the other models. The resulting histogram is presented in 
Fig.~\ref{totalmasshist}. The broad variation of peaks in the individual
histograms seen in Fig.~\ref{masshist} leads to a smooth power-law
tail in the high mass end of the composite histogram. Fig.~\ref{totalmasshist}
reveals a slope $d\log N/d\log M \approx -2$ in this region, only somewhat
steeper than
the value $d\log N/d\log M \approx -1.5$ measured for example by \citet{Mott}.
We note that this general mechanism, arising from an initially inhomogeneous
distribution of mass-to-flux ratio, is an interesting new possibility for 
explaining the observed broad core mass distributions. 
An important point is that a relatively {\it narrow} distribution of
initial mass-to-flux ratios (factor of a few variation) can lead to a relatively
broad distribution of fragment masses.
This mechanism remains to be explored
more generally and does not exclude the
occurrence of other mechanisms like
competitive accretion in a more global model \citep{Bonn}, a temporal 
spread of core accretion lifetimes \citep{Myer00,BJ04}, and 
turbulent fragmentation \citep[e.g.][]{Pado, Kless, Gamm,Till}. 


The richness of the physics revealed by our models of gravitational fragmentation
up to runaway collapse of the first core, under conditions of varying mass-to-flux ratio, 
ionization level, and external pressure, pave the way for more extensive models 
in the future. 
Adding the effects of initial cloud turbulence, implementing a technique for integrating past the
formation of the first generation of stars, and including some form of energy 
feedback from star formation, remain to be done.
The addition of new and more complex effects will be facilitated by the fact that
the thin-sheet approximation allows efficient calculation of the fragmentation
process while retaining a high level of realism. In this context we point out that
the recent fully three-dimensional fragmentation simulation of 
\citet{Kudo07} with magnetic fields and ambipolar diffusion bears out the 
main physical results presented by BC04.

\section{Summary}
\label{s:summ}

We have carried out a large number of model simulations to study the 
effect of initial mass-to-flux ratio, neutral-ion coupling, and external
pressure on dense core formation from gravitational 
fragmentation of isothermal sheet-like layers that may be embedded within larger
molecular cloud envelopes. 
Our simulation box is periodic in the lateral ($x,y)$ directions and 
typically span four nonmagnetic (Jeans) fragmentation scales in each of these
directions.
The simulations reveal a wide range of outcomes, from
the unique transcritical fragmentation mode into massive cores, to the 
pressure-driven fragmentation into dense clusters. 
We emphasize the following main results of the paper:

\begin{enumerate}

\item{\it Fragmentation Spacing.}
The average spacings of nonlinearly developed fragments are generally
in excellent agreement 
with the preferred fragmentation scale of linear perturbation theory 
\citep{CB06}, although there is
definite irregularity in any simulation, with variation of fragment spacings.
Both significantly subcritical {\it and} highly supercritical clouds 
have average fragmentation scales $\lamavg \approx 2 \pi Z_0$, where   
$Z_0$ is the half-thickness of the background state. 
The transcritical ($\mui \approx 1$) models exhibit very large (super-Jeans) 
average fragment 
spacings, although there is evidence for nonlinear second-stage fragmentation in 
some cases, which makes the average spacing slightly smaller than predicted by 
linear theory.  
Variation of the ionization fraction by a factor of two above and below the standard
value does not have a big effect on fragment spacing. However, an external pressure
dominated sheet undergoes dramatically smaller scale fragmentation to form 
a dense cluster.


\item{\it Time Evolution to Runaway.} The times $t_{\rm run}$ for various models to reach
runaway collapse of the first core varies significantly for models with differing
initial dimensionless mass-to-flux ratio $\mui$, neutral-ion coupling
parameter $\tniitil$, and dimensionless external pressure $\Pexttil$. 
Values of $t_{\rm run}$ range from 0.45 Myr to 9.53 Myr, each scaling as 
$(\Nni/10^{22}\,\cms)^{-1} (T/10\K)^{1/2}$.
The supercritical
clouds evolve much more rapidly than the critical or subcritical clouds, with 
the highly supercritical clouds evolving $\approx 10$ times more rapidly than a
critical cloud, for the typical level of neutral-ion coupling. A critical cloud 
in turn evolves more rapidly than a subcritical cloud, but the variation is a factor
of order unity for plausible initial values of $\mui$; for example the 
$\mui = 0.5$ model reaches runaway collapse in a time that is 1.7 times longer than for the
$\mui=1$ model. 
In all cases, the time to runaway collapse is $\approx 10 \, \taugm$ when starting with
small-amplitude white noise perturbations, where
$\taugm$ is the growth time of the fastest growing eigenmode mode in linear
perturbation theory. 
The quantity $\taugm$ itself varies from $\approx Z_0/\cs$ for highly 
supercritical 
models to $\approx 10 Z_0/\cs$ for highly subcritical models (for a typical neutral-ion
coupling level).
The effect of varying $\tniitil$ (and hence the initial ionization fraction $x_{\rm i,0}$)
is that $t_{\rm run} \propto \tniitil^{-1} \propto x_{\rm i,0}$ approximately for ambipolar-drift-driven (critical or subcritical) fragmentation, so that the canonical $\taugm \approx 10 Z_0/\cs$
and our calculated $t_{\rm run} \approx 100 Z_0/\cs$ are both possibly subject to significant variation.
A pressure dominated cloud with $\Pexttil=10$ has $t_{\rm run}$ about 5-6 times shorter
than clouds with small external pressure but other parameters held fixed.


\item{\it Velocities in the Nonlinear Regime.}
Maximum infall speeds of neutrals can become supersonic on core scales in the 
supercritical clouds, but remain subsonic for critical or subcritical clouds. 
The latter is true even if the ionization fraction is reduced by a factor of
two.
The ion speeds in the cores closely follow the neutral speeds but are somewhat smaller,
since the gravitationally-driven motion of the neutrals is generally opposed in
the plane of the sheet by magnetic fields. 

\item{\it Core Shapes.}
An extensive compilation of core shapes shows that the distributions have a
peak that is near-circular for the subcritical and critical fragmentation models. 
However, the cores become more elongated in the sheet for supercritical clouds,
with a mean axis ratio in the sheet plane $\approx 0.5$. The half-thickness remains
smaller than either semiminor or semimajor axis in the sheet, so that the cores
are triaxial and preferentially flattened along the direction of the mean 
magnetic field.
Preliminary comparison of our results with published deprojections
of the observed axis ratios of cores shows the best agreement with our 
subcritical models.

\item{\it Core Mass Distributions.}
An extensive compilation of core masses shows that the peak of the distributions 
are related to the preferred fragmentation mass $\mgm$ of linear theory.
Transcritical fragmentation yields the largest peak masses while the highly
supercritical and decidely subcritical limits have smaller (Jeans-like) peaks
and similar distributions. That the peak mass is strongly selected is seen in
the very sharp drops in the distributions for greater masses. This means that
the observed relatively broad-tailed core mass distributions cannot be explained by
a pure local gravitational fragmentation process in a medium of uniform background
column density and mass-to-flux ratio. 
Cores defined as significantly supercritical regions within a larger subcritical
common envelope also have a very narrowly-peaked distribution. 
However, a composite mass histogram that may mimic the effect of an 
inhomogeneous assortment of 
initial mass-to-flux ratios,
does produce a broad core mass distribution that resembles its observed counterparts.

\item{\it Magnetic Field Line Structure.}
Contraction of cores within a supercritical cloud yields significant
curvature of the magnetic field lines and very apparent hourglass morphologies,
since contraction proceeds primarily with field-line dragging. The transcritical
and subcritical clouds form cores through a process driven in large part by
neutral-ion slip, and result in lesser curvature in the magnetic field. 
This holds out the hope of using the observed curvature of 
possible future observations of hourglass magnetic fields on the core scale 
as a proxy for measuring the ambient mass-to-flux ratio.

\end{enumerate}


\section*{Acknowledgements}

We thank the anonymous referee for comments which significantly improved
the manuscript. We also thank
Wolfgang Dapp for valuable comments on the manuscript and thank 
both him and Stephanie Keating for creating 
several color images and animations. 
The IFRIT package, developed by Nick Gnedin, was used fruitfully 
to create some color images and magnetic field line visualizations.
Our IDL code benefited from the use of an Adams-Bashforth-Moulton
ODE solver converted to IDL format by Craig Markwardt from the public 
domain Fortran routine written by
L. F. Shampine and H. A. Watts of Sandia Laboratories.
SB was supported by a grant from the Natural Sciences and Engineering
Research Council (NSERC) of Canada. 
JW was supported by an NSERC Undergraduate Summer Research Award. 
SB would also like to thank the KITP Santa Barbara
for their hospitality during the final stages of writing this paper,
when this research was supported in part by the National Science 
Foundation under Grant No. NSF PHY05-51164.

\end{document}